\documentclass[twocolumn]{aastex63}
\usepackage{graphicx}
\usepackage{subfigure}
\usepackage{color, hyperref, epsfig}
\usepackage{apjfonts, natbib}
\usepackage{appendix}
\usepackage{amsmath}
\usepackage{float}
\usepackage{bm}

\newcommand{\lum}{erg~s\ensuremath{^{-1}}}

\shorttitle{ASASSN-18ap}
\shortauthors{Wang et al.} 

\submitjournal{\apj}
\received{ 2023 June 13}
\revised{2023 December 18}
\accepted{2024 February 11}

\begin{document}
\title{ASASSN-18ap: A Dusty Tidal Disruption Event Candidate with an Early Bump in the Light Curve }
\author[0000-0003-4225-5442]{Yibo~Wang}
\affiliation{CAS Key laboratory for Research in Galaxies and Cosmology,
Department of Astronomy, University of Science and Technology of China, 
Hefei, 230026, China; wybustc@mail.ustc.edu.cn,twang@ustc.edu.cn,jnac@ustc.edu.cn}
\affiliation{School of Astronomy and Space Sciences,
University of Science and Technology of China, Hefei, 230026, China}

\author[0000-0002-1517-6792]{Tinggui Wang}
\affiliation{CAS Key laboratory for Research in Galaxies and Cosmology,
Department of Astronomy, University of Science and Technology of China, 
Hefei, 230026, China; wybustc@mail.ustc.edu.cn,twang@ustc.edu.cn,jnac@ustc.edu.cn}
\affiliation{School of Astronomy and Space Sciences,
University of Science and Technology of China, Hefei, 230026, China}

\author[0000-0002-7152-3621]{Ning Jiang}
\affiliation{CAS Key laboratory for Research in Galaxies and Cosmology,
Department of Astronomy, University of Science and Technology of China, 
Hefei, 230026, China; wybustc@mail.ustc.edu.cn,twang@ustc.edu.cn,jnac@ustc.edu.cn}
\affiliation{School of Astronomy and Space Sciences,
University of Science and Technology of China, Hefei, 230026, China}

\author[0000-0002-1542-8080]{Xiaer Zhang}
\affiliation{CAS Key laboratory for Research in Galaxies and Cosmology,
Department of Astronomy, University of Science and Technology of China, 
Hefei, 230026, China; wybustc@mail.ustc.edu.cn,twang@ustc.edu.cn,jnac@ustc.edu.cn}
\affiliation{School of Astronomy and Space Sciences,
University of Science and Technology of China, Hefei, 230026, China}

\author[0000-0003-3824-9496]{Jiazheng Zhu}
\affiliation{CAS Key laboratory for Research in Galaxies and Cosmology,
Department of Astronomy, University of Science and Technology of China, 
Hefei, 230026, China; wybustc@mail.ustc.edu.cn,twang@ustc.edu.cn,jnac@ustc.edu.cn}
\affiliation{School of Astronomy and Space Sciences,
University of Science and Technology of China, Hefei, 230026, China}

\author[0000-0002-7020-4290]{Xinwen~Shu}
\affiliation{Department of Physics, Anhui Normal University, Wuhu, Anhui, 241000, People's Republic of China}

\author[0000-0001-7689-6382]{Shifeng Huang}
\affiliation{CAS Key laboratory for Research in Galaxies and Cosmology,
Department of Astronomy, University of Science and Technology of China, 
Hefei, 230026, China; wybustc@mail.ustc.edu.cn,twang@ustc.edu.cn,jnac@ustc.edu.cn}
\affiliation{School of Astronomy and Space Sciences,
University of Science and Technology of China, Hefei, 230026, China}

\author{FaBao~Zhang}
\affiliation{Department of Physics, Anhui Normal University, Wuhu, Anhui, 241000, People's Republic of China}

\author[0000-0001-6938-8670]{Zhenfeng~Sheng}
\affiliation{Institute of Deep Space Sciences, Deep Space Exploration Laboratory, Hefei 230026, China}
\affiliation{School of Astronomy and Space Sciences,
University of Science and Technology of China, Hefei, 230026, China}

\author[0000-0003-4959-1625]{Zheyu~Lin}
\affiliation{CAS Key laboratory for Research in Galaxies and Cosmology,
Department of Astronomy, University of Science and Technology of China, 
Hefei, 230026, China; wybustc@mail.ustc.edu.cn,twang@ustc.edu.cn,jnac@ustc.edu.cn}
\affiliation{School of Astronomy and Space Sciences,
University of Science and Technology of China, Hefei, 230026, China}

\begin{abstract}
We re-examined the classification of the optical transient ASASSN-18ap, which was initially identified as a supernova (SNe) upon its discovery. Based on newly emerged phenomena, such as a delayed luminous infrared outburst and the emergence of luminous coronal emission lines, we suggest that ASASSN-18ap is more likely a tidal disruption event (TDE) in a dusty environment, rather than a supernova. The total energy in the infrared outburst is $\rm 3.1\times10^{51}$ erg, which is an order of magnitude higher than the total energy in the optical-to-ultraviolet range, indicating a large dust extinction, an extra-EUV component, or anisotropic continuum emission. A bumpy feature appeared in the optical light curve at the start of brightening, which was reported in a couple of TDEs very recently. This early bump may have been overlooked in the past due to the lack of sufficient sampling of the light curves of most TDEs during their ascending phase, and it could provide insight into the origin of optical emission.
\end{abstract}

\section{introduction} 
\label{intro}
Tidal disruption events (TDEs) occur when a star approaches a supermassive black hole (SMBH) close enough to be pulled apart, resulting in a flare of electromagnetic radiation that peaks in the ultraviolet (UV) to soft X-ray band as part of the debris is accreted by the SMBH (e.g., \citealt{Rees1988,EK1989,SQ2009,LR2011,Gezari2021}). TDEs were first theorized in the 1970s (e.g., \citealt{Hills1975}), and the earliest candidates were discovered in ROSAT archival data two decades later (e.g., \citealt{Bade1996,Komossa1999}). TDEs provide an excellent opportunity to explore dormant SMBHs in galaxy centers and to study the evolution of supermassive black-hole accretion systems on short timescales of about a year. However, the occurrence rate of such events is estimated to be about $10^{-5}-10^{-4}\,$galaxy$^{-1}\,$year$^{-1}$ (e.g., \citealt{Wang2004,Stone2016,vV2018,Stone2020}). In the past two decades, about 100 TDE candidates have been discovered in multiband surveys, mostly in X-ray and optical wavelengths (e.g., \citealt{Gezari2021,Sazonov2021,Hammerstein2023}). Although the first discovery was made more than a decade after X-ray one (\citealt{vV2011}), optical TDEs become the dominant population, benefiting from facilities devoted to wide-field and fast-sky optical surveys, such as All-Sky Automated Survey for Supernovae (ASAS-SN; \citealt{Shappee2014,Kochanek2017}), the Panoramic Survey Telescope \& Rapid Response System (Pan-STARRS or PS1; \citealt{PS1}), the intermediate Palomar Transient Factory (iPTF; \citealt{Kulkarni2013}), the Asteroid Terrestrial Impact Last Alert System (ATLAS;  \citealt{ATLAS,ATLASforTransient}), and the Zwicky Transient Facility (ZTF; \citealt{Bellm2019}). These events exhibit diverse properties in their light curves, emission lines and X-ray (e.g., \citealt{Saxton2020,vV2020ssr,vV2021ApJ, Gezari2021}). Most optical TDEs have not been detected in the X-ray band, with several exceptions, such as ASASSN-14li (\citealt{Holoien2016}), ASASSN-15oi (\citealt{Gezari2017}), AT2018fyk (\citealt{Wevers2021,Wevers2023}), AT2019qiz (\citealt{Nicholl2020}), and AT2017gge 
 (\citealt{Onori2022,Wang2022ApJL}). Therefore, despite the considerable efforts made in theories and simulations of these events (e.g., \citealt{Lodato2020,Roth2020}), understanding the essence of TDEs remains a challenge.

Optical TDEs usually reach a maximum luminosity of around $\rm 10^{43.3}-10^{45.8} erg/s$ over the course of a month, followed by a power-law decrease that can last for years (e.g., \citealt{vV2020ssr,Lin2022ApJ,Hammerstein2023}). It is thought that this decline is due to mass fallback rates, although this is still a subject of discussion (e.g., \citealt{Gezari2021}). The optical-to-ultraviolet (OUV) spectral energy distribution (SED) can be usually represented by a black body with a temperature of approximately 15,000 to 40,000 Kelvin, which can remain stable for weeks or months. (e.g., \citealt{vV2020ssr,Hammerstein2023}). The constancy of the temperature is useful in distinguishing optical TDEs from imposters. Optical TDEs are spectroscopicaly characterized by broad emission lines of H Balmer or/and $\rm He\,\textsc{ii}$ on the top of a strong blue continuum (e.g., \citealt{Arcavi2014,vV2020ssr,vV2021ApJ,Charalampopoulos2022,Hammerstein2023}). \cite{vV2021ApJ} classified TDEs into three categories based on the existence or absence of these emission lines: TDE-H (only H Balmer lines detected), TDE-He ($\rm He\,\textsc{ii}$ only) and TDE-H+He (both H and $\rm He\,\textsc{ii}$). 
The spectral classification of TDEs is even more complex, with the identification of broad Bowen fluorescent lines (including $\rm N\,\textsc{iii}\lambda4640$, $\rm N\,\textsc{iii}\lambda4100$, $\rm O\,\textsc{iii}\lambda3760$) in iPTF15af  (\citealt{Blagorodnova2019}), ASASSN-14li, iPTF16axa, AT2018dyb (\citealt{Leloudas2019}), and iPTF16fnl (\citealt{Onori2019}). This forms a subclass of N-rich TDEs, referred to as TDE-Bowen. Interestingly, \cite{Leloudas2019} found that almost all TDE-H+He show these Bowen features, which was further confirmed by the TDEs discovered by the ZTF survey (e.g., \citealt{vV2021ApJ,Hammerstein2023}). Furthermore, \citealt{vV2021ApJ} demonstrated that TDE-Bowen has a smaller radius and a higher temperature of blackbody, which are beneficial to powering Bowen fluorescence. In fact, the Bowen process was activated by the ionization (requiring photons with energies greater than 54 eV) and recombination of $\rm He\,\textsc{ii}$ (\citealt{Bowen1934,Bowen1935}), providing observational proof of the presence of obscured and reprocessed EUV/X-ray emission in optical TDEs. (e.g., \citealt{Leloudas2019}). The profiles of the emission lines mentioned above usually are very broad with FWHM  $\rm \sim 10000\,km/s$ and in general pure emission line profile without absorption or P-cygni feature (e.g., \citealt{Arcavi2014,vV2020ssr,Gezari2021}). Some events display double-peaked or boxy profile, such as PTF09dj (\citealt{Arcavi2014}), AT2018zr (\citealt{Holoien2019}) and AT2018hyz (\citealt{Hung2020,Short2020}). A more common trend is that the emission lines profile would become narrower at late-time, which is opposite to the reverberation of fading continuum in AGN (e.g., \citealt{Holoien2016}), in which the kinematic of emitting gas dominates the spectral line width. However, \cite{RothKasen2018} shows that the emission lines width in some TDEs  may be set by the electron scattering optical depth rather than gas kinematics through the radiative transfer calculations.  Notably, the recent work by \cite{Hammerstein2023} carefully studied a sample of 30 TDEs discovered by ZTF and reported a new spectroscopic class, termed TDE-featureless. These TDE-featureless are characterized by a blue continuum and featureless spectra, along with larger bolometric luminosities, blackbody temperatures, and blackbody radii at peak compared to the common TDE population.

However, it has been suggested that optical surveys may be strongly biased against a population of highly dust enshrouded TDEs (\citealt{Jiang2021ApJ,Roth2021,Reynolds2022}). These dusty events can be revealed through their reprocessed infrared emission, called IR echoes. This has been validated theoretically and observationally (\citealt{Lu2016,Jiang2016,vV2016,Mattila2018}). Using the database of the Wide-field Infrared Survey Explorer (WISE; \citealt{Wright2010}) and its successor, the Near-Earth Object WISE Reactivation mission (NEOWISE-R; \citealt{Mainzer2014}), \cite{Jiang2021ApJS} has constructed a large sample of mid-infrared outbursts in nearby galaxies (MIRONG) to statistically search for transients in infrared bands. Subsequent spectroscopic follow-up has found 14 TDE candidates (\citealt{Wang2022ApJS}), including a robust dusty TDE, ATLAS17jrp/AT2017gge (\citealt{Onori2022,Wang2022ApJL}). 

In this paper, we focus on another promising dusty TDE candidate, ASASSN-18ap, which was first reported by the ASASSN ( \citealt{ASASSN18ap-report}) in the host galaxy SDSSJ014642.44+323029.3 at a redshift of 0.037503 (\citealt{Springob2005}). Initially, it has been classified as a supernova candidate (called SN2018gn) by \cite{ASASSN18ap-class} based on an early optical spectrum. Interestingly, \cite{SNeI-IR} found it has a distinctly high peak infrared luminosity compared to other supernovae and thus proposed a TDE scenario. In this work, we discuss how the light curves and spectroscopic features of ASASSN-18ap can be consistent with the TDE scenario and carefully exclude the SN scenario.

The paper is structured as follows. In Section \ref{sec2}, we describe the data reduction and preliminary analysis of multi-band observations from radio to X-ray for ASASSN-18ap. The resulting multi-wavelength light curves and optical spectra are analyzed in Section \ref{sec3}. In Section \ref{sec4}, we discuss the possible nature of ASASSN-18ap and we also highlight the high infrared luminosity of ASASSN-18ap and its bump feature at the onset of brightening. Finally, we present our conclusion in Section \ref{sec5}. We assume a cosmology with $\rm H_{0} =70$ km~s$^{-1}$~Mpc$^{-1}$, $\rm \Omega_{m} = 0.3$, and $\rm \Omega_{\Lambda} = 0.7$ throughout the article.

\section{observation and data reduction }
\label{sec2} 
\subsection{Optical photometry} 
\label{sec2.1}
We first checked the archival optical light curves of ASASSN-18ap from various surveys, including ASAS-SN (\citealt{Shappee2014,Kochanek2017}), ATLAS ( \citealt{ATLAS,ATLASforTransient}), Gaia Alerts (\cite{GaiaAlerts}), and ZTF (\citealt{Bellm2019}). We processed the light curves as follows. Due to their poor quality and overlap with the ATLAS light curves, we excluded the ASAS-SN~\footnote{https://asas-sn.osu.edu/} light curves in subsequent analyses. Furthermore, the ASASSN-18ap has been monitored multiple times by the Swift Ultra-Violet/Optical Telescope (Swift/UVOT; \citealt{Swift/UVOT}), and the data reduction steps for this data-set are described in Section \ref{sec2.3}.

\emph{ATLAS photometry}.
We first obtained the point-spread-function (PSF) profile fitting photometry from the ATLAS Forced Photometry Website~\footnote{https://fallingstar-data.com/forcedphot/}. This forced photometry requires the PSF profile fit to be applied at the coordinates specified by the user on the differential image. For the retrieved ATLAS $o$ and $c$ band light curves, we first filtered out exposures with a sky background brighter than 19 magnitude~\footnote{The limiting magnitude of ATLAS is about 19-19.5(\citealt{ATLASforTransient}), and thus measurements with sky background brighter than 19 magnitude usually are unreliable. This filter is effective in eliminating outliers and preserving most reliable measurements.} and manually eliminated remaining outliers. We then binned the light curves every half-day to improve the signal-to-noise ratio. Due to some schematics, the baseline level, i.e., the flux before the transient, is around a negative value, which, however, should be zero. We then used the average flux before the outburst ($\rm 57950\,<\,MJD\,<\,58110$) to correct the baseline to the zero level. 

\emph{Gaia photometry}. 
The Gaia G-band light curve was retrieved from the Gaia Photometric Science Alerts Website~\footnote{http://gsaweb.ast.cam.ac.uk/alerts/alert/Gaia18apd/}, in which the light curve is extracted by means of a PSF/LSF (point/line spread function, \citealt{GaiaAlerts}) fitting. Since the photometry was not host-subtracted, we derived the outburst flux by subtracting the average flux before the outburst (MJD $<$ 58115) from the light curve. We then binned the data points around the peak (58189$<$MJD$<$58195) every day and binned the remaining data points every 60 days.

\emph{ZTF photometry}. 
We acquired the ZTF light curves of ASASSN-18ap through the ZTF forced photometry service~\footnote{https://ztfweb.ipac.caltech.edu/cgi-bin/requestForcedPhotometry.cgi} (\citealt{Masci2019}). This service performs the PSF profile fitting at a user-specified location on archived difference images from the ZTF survey. We then filter out photometric data that were affected by bad pixels or bad seeing. Since the reference image contained the transient flux, we calculated the average flux of post-transient exposures (59800$\rm < MJD<$59873), in which the transient flux had almost declined to zero, and subtracted the flux from all the data subsequently.

\emph{LCOGT photometry}.
We also collected the archival data of ASASSN-18gn from the Las Cumbres Observatory Global Telescope network~\footnote{https://archive.lco.global/} (LCOGT; \citealt{LCOGT}), and found three epochs of observations. The r- and i-band LCOGT images were resampled and aligned to PanSTARRS (PS1) stack images, followed by image subtraction using {\tt HOTPANTS} \citep{Becker2015}. The photometry results were calibrated to PS1 system \citep{Tonry2012}. Since the only three epochs of observations basically overlapped with the Swift/UVOT photometry, we did not include the LCO results for the analysis below in this work, except for the discussion in Appendix \ref{appendiceC}.
 
The middle panel of Figure \ref{SN2018gnlc} shows the optical light curves from several telescopes, including ATLAS-$o$, Swift-$U$, Gaia-$G$, and ZTF-$r$, which were combined to create a complete optical light curve for ASASSN-18ap. We took the date when the ATLAS-$o$ band reached its maximum flux ($\rm MJD\,58161$) as the optical peak position and found that it had a rising timescale to the peak of about 40 days. Interestingly, there is a bump in the ATLAS-$o$ band light curve at the onset of the rising phase (see Figure \ref{SN2018gnlc} and details will be described in Section \ref{sec4.6}).


\subsection{Archival WISE photometry}
\label{sec2.2}
The NEOWISE-R survey provided multiple mid-infrared photometry (56681.4-59790.68) for ASASSN-18ap. We obtained single-exposure photometry in the W1 (centered at 3.4 $\rm \mu m$) and W2 (4.6 $\rm \mu m$) bands from the public NEOWISE-R Single Exposure (L1b) Source Table~\footnote{https://irsa.ipac.caltech.edu/cgi-bin/Gator/nph-scan?projshort=WISE\&mission=irsa}, which is measured by PSF profile fitting. We removed bad data points from the retrieved single-exposure data based on the following quality flags: poor quality frames ($qi\_fact<1$), charged particle hits ($saa\_sep<5$), scattered moonlight ($moon\_masked=1$), artifacts ($cc\_flags \neq0$), and multiple PSF components ($nb>1$ and $na>0$). We then binned the single-exposure photometry every six months and used the average flux before $\rm MJD\,58000$ as a reference to remove the host-galaxy contribution. The binned host-subtracted fluxes are plotted in the middle panel of Figure \ref{SN2018gnlc}. The mid-infrared light curves took about 600 days to reach their peak at $\rm MJD\,58695$, which was the most luminous point in the W1 band and also in blackbody luminosity. This peak was about 533 days delayed compared to the peak of the optical light curves.
 
\subsection{Swift observation } 
\label{sec2.3}
\begin{figure*}[htb]
\figurenum{1}
\centering
\begin{minipage}{0.9\textwidth}
\centering{\includegraphics[angle=0,width=1.0\textwidth]{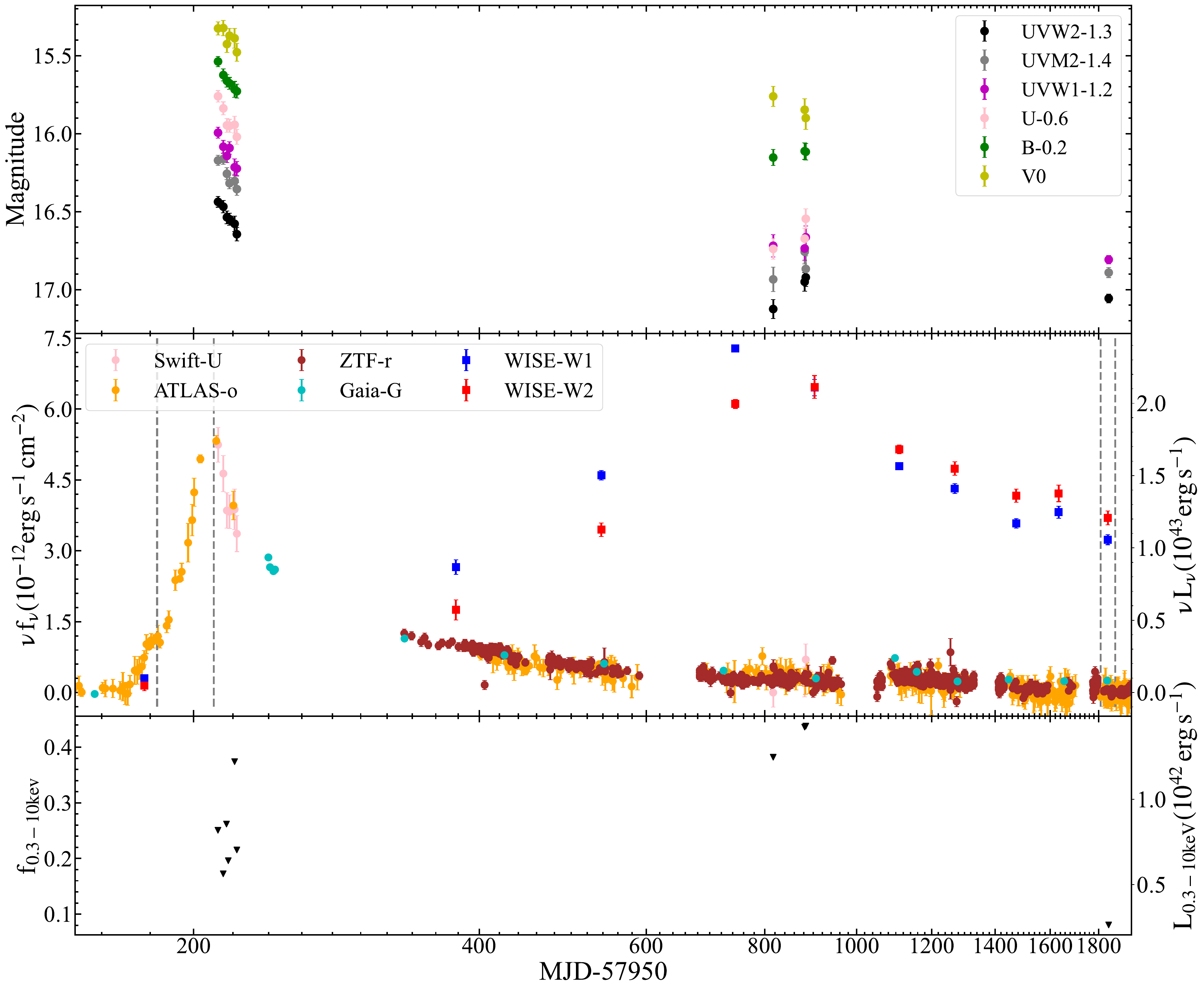}}
\end{minipage}
\caption{The infrared to X-ray light curves of the ASASSN-18ap. (a) In the top panel, we show light curves from the Swift/UVOT. (b) The middle panel displays the two host-subtracted infrared light curves (WISE-W1, WISE-W2) and four host-subtracted optical light curves (ATLAS-o, Swift/U, ZTF-r, Gaia-G), which were spliced into a complete optical light curve naturally. The dashed gray line indicates the dates of the spectroscopic observations. (c) The upper limits of the unabsorbed X-ray flux, converted from the upper limits of the count rates from Swift/XRT observations assuming a $\rm kT\sim50\,eV$ blackbody SED shape for a typical TDE, are shown in the bottom panel. All the photometry shown in the top two panels of this figure is available as the data behind the figure. \label{SN2018gnlc} }
\end{figure*}

ASASSN-18ap was observed six times between MJD 58162-58172 (PI: Dong), three times between MJD 58766-58834 (PI: Brown), and once more recently at MJD 59794 (PI: Wang) (see Table~\ref{swift/XRT}). The Swift/UVOT photometry was measured using 5" apertures with the UVOTSOURCE task in the HEASoft v6.29 package, and the AB magnitude was calibrated in the Swift photometric system. The top panel of Figure \ref{SN2018gnlc} shows the measured photometry. We used the latest observation as the reference for the three UV bands to remove the host galaxy contribution, considering its adequate exposure time and high signal-to-noise ratio. Meanwhile, the $U$, $B$, and $V$ bands were subtracted by the faintest one at $\rm MJD\,58767$.

For Swift/XRT (\citealt{Swift/XRT}) observations, we used XRTPIPELINE in HEASOFT 6.30.1 with the most recent calibration files available at the time to reproduce the event files. Then, using the task XRTPRODUCTS, we extracted the source with a circular region of radius 20" and the background from an annulus with an inner radius of 60" and an outer radius of 200". None of the detections were significant enough to reach 99.7\% confidence according to the Poisson distribution, including the two stacked images from adjacent observations.  Based on the Bayesian method in \cite{Kraft1991}, we calculated the upper limits of the count rates at the 99.7\% confidence level and list them in Table \ref{swift/XRT}, along with the extracted source counts and background levels.  Using the {\tt PIMMS}~\footnote{https://heasarc.gsfc.nasa.gov/cgi-bin/Tools/w3pimms/w3pimms.pl} tool, we converted the count rate upper limits to flux without considering any intrinsic absorption from the host, assuming a $\rm kT\sim50\,eV$ blackbody SED shape for a typical TDE (e.g., \citealt{Saxton2020,Gezari2021}) or the APEC model with $\rm kT\sim1\,keV$ to describe the emission of high temperature plasma for CSM interaction of SNe IIn (\citealt{Smith2001, Katsuda2014, Chandra2015}). We considered Galactic absorption with a column density of $\rm 3.86\times10^{20}\,cm^{-2}$ (\citealt{HI4PICollaboration2016}). The resulting unabsorbed fluxes are listed in Table \ref{swift/XRT} and are plotted in the bottom panel of Figure \ref{SN2018gnlc}.

\subsection{Radio observations} 
\label{sec2.4}
\begin{figure}[htb]
\figurenum{2}
\centering
\begin{minipage}{0.45\textwidth}
\centering{\includegraphics[angle=0,width=1.0\textwidth]{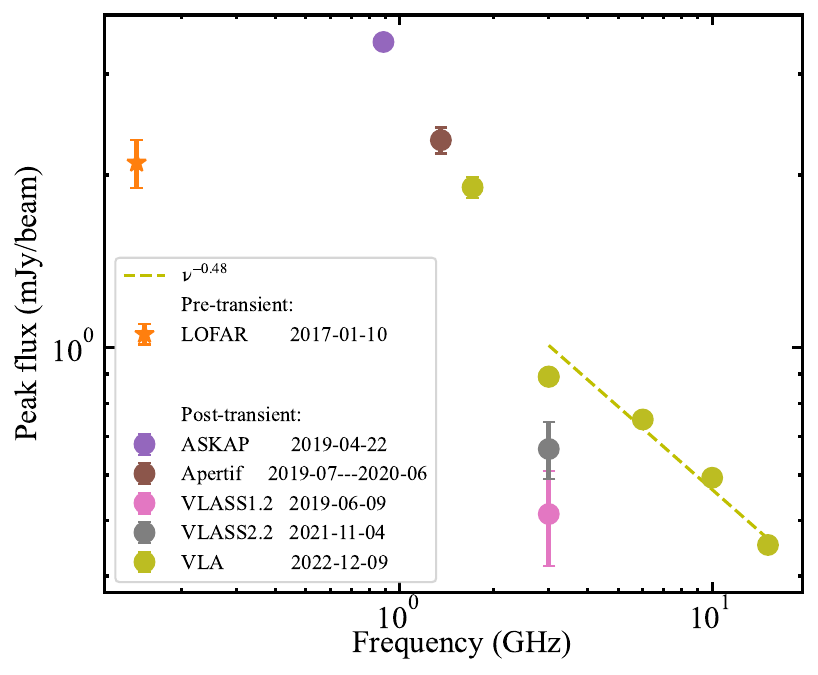}}
\end{minipage}
\caption{We plot the peak flux as a function of frequency for all radio detections of ASASSN-18ap. The pre-transient LOFAR detection is shown as orange five-pointed star, and the post-transient radio detections are shown as points with different colors indicating different observation time and instruments. The yellow dashed line shows the powerlaw fitting to the 3-15GHz data of the recent VLA broadband spectra (yellow points). The data of VLA spectra and other archival radio detection are available as Data behind Figure. \label{radio-plot}}
\end{figure}

Prior to the occurrence of ASASSN-18ap, its host galaxy was detected in the Low-Frequency Array (LOFAR) Two-meter Sky Survey (LoTSS; \citealt{LoTSSDR2}) at 0.144 GHz, but no other data were found. After the outburst, ASASSN-18ap was covered by the Rapid ASKAP Continuum Survey (RACS; \citealt{RACS}) using the Australian Square Kilometre Array Pathfinder (ASKAP) at 464 days since the optical peak, and was also detected by the Aperture Tile In Focus (Apertif; \citealt{Apertif}). The source was not significantly detected in the Very Large Array (VLA) Sky Survey (VLASS;\citealt{VLASS} ) at epoch 1.2 (482 days after the optical peak), and was marginally detected at epoch 2.2 (1361 days) with a flux consistent with epoch 1.1 within their uncertainty. Additionally, we proposed a multi-frequency observation with VLA (Program ID 22B-319; PI, Wang) in the C array configuration and detected the source in the five bands L, S, C, X, Ku. We processed the VLA data in CASA v5.3.0 following the standard data reduction procedure described in \cite{CASA2022}. We plot them together with the above archival radio observations in Figure \ref{radio-plot}. The source appears extended at frequencies 3~GHz or lower. However, because of the lack of multi-frequency spectrum before the outburst of ASASSN-18ap, we could not determine the host-galaxy contribution for the broad-band VLA spectra we acquired, and, in fact, the host galaxy can account for a significant fraction of the radio emission. According to Equation (6) of \cite{WISESFR}, we first estimated that the present star formation rate (SFR) is $\rm \sim 5\ M_\odot year^{-1}$ from the W4 photometry of the host (see Table \ref{host-photos}). Then, using the SFR-radio continuum relation in \cite{radioSFR}, we calculated the 1.4 GHz radio luminosity for the host galaxy as about $\rm 8.54\times10^{28}~erg\,s^{-1}\,Hz^{-1}$, which converted to an observed flux density of 2.6 mJy, comparable to the VLA observations. Additionally, radio emission in 3-15 GHz can be described as a steep spectrum with $\alpha=-0.48$ ($S_{\nu}={\nu^{\alpha}}$), which is consistent with that of star-forming galaxies (e.g.,\citealt{Gioia1982}), indicating no obvious radio brightening since the optical outburst, at least at the time of radio observations. The radio emission at lower frequencies has a steeper slope, which is likely due to observations with different (worse) resolutions. The enhancement of the 3 GHZ flux of VLA compared to the VLASS may also be due to a lower resolution. Given that a significant contribution from the host galaxy was very likely and it was difficult to eliminate this component, we abandoned a detailed analysis of the radio spectra. 

\subsection{Spectroscopic observation and data reduction} 
\label{sec2.5}
\begin{figure*}[htb]
\figurenum{3}
\centering
\begin{minipage}{0.9\textwidth}
\centering{\includegraphics[angle=0,width=1.0\textwidth]{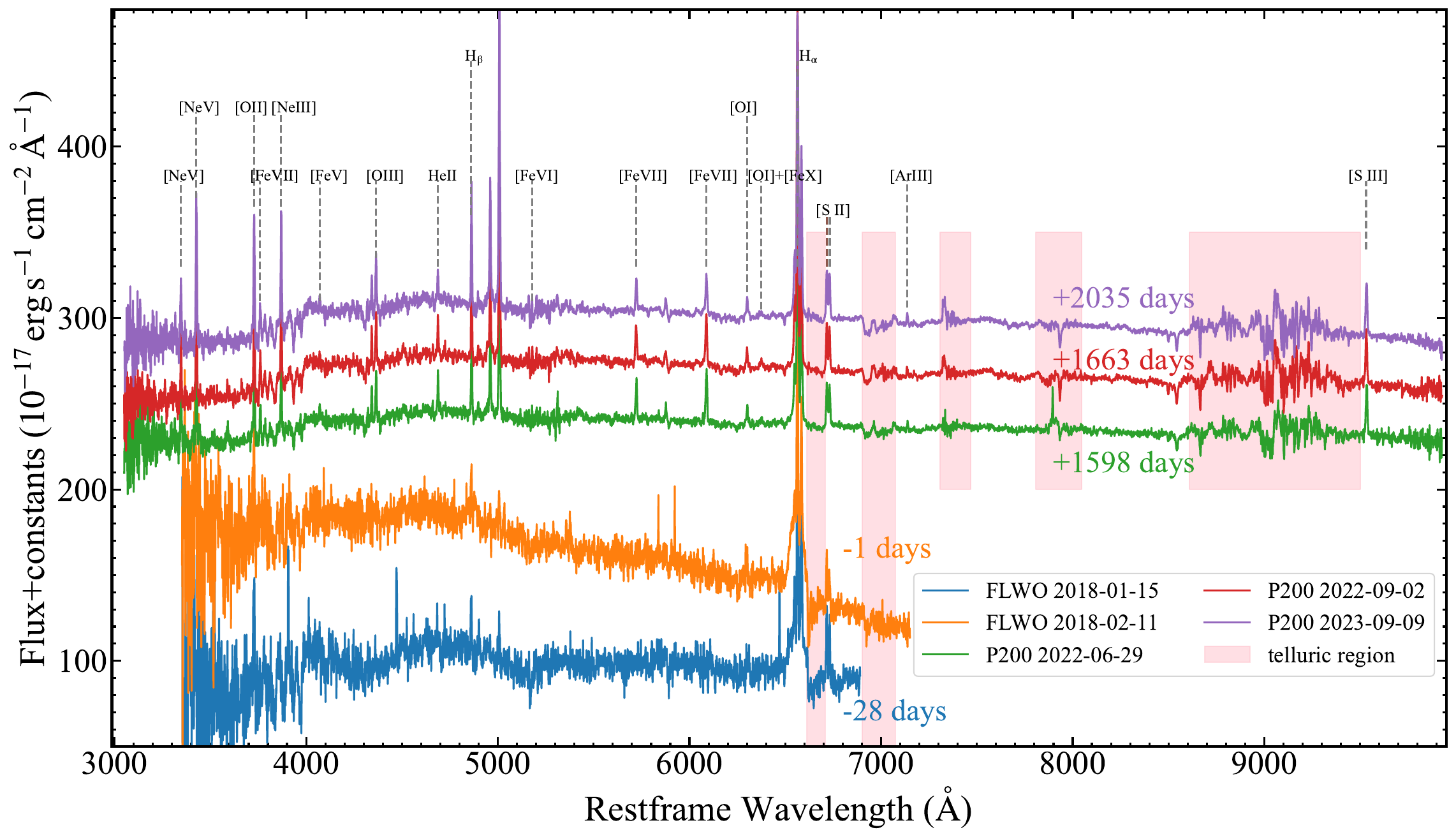}}
\end{minipage}
\caption{Five spectra acquired for ASASSN-18ap are shown in the figure. The pink shadow indicates the telluric region,  and the vertical dashed lines mark some of the characteristic emission lines. The phases relative to the peak are displayed on the right side of each spectrum with the same color. \label{all_spec}}
\end{figure*}

Immediately after the discovery of ASASSN-18ap, \cite{ASASSN18ap-class} acquired an optical spectrum on UT2018-01-15, around the early bump of the optical light curve, using the FAST Spectrograph on the Fred L. Whipple Observatory 1.5 m Tillinghast telescope (FLWO/FAST; \citealt{Fabricant1998}). We retrieved this spectrum from the TNS website~\footnote{https://www.wis-tns.org/object/2018gn}. Searching in the CfA Optical/Infrared Science Archive~\footnote{https://oirsa.cfa.harvard.edu/search/} (\citealt{Mink2021}), we found another spectrum taken on UT2018-02-11 (near the peak of the optical light curve) with FLWO/FAST, which has not been reported before. Flux calibration has not been done for the archival reduced spectrum, and we did this using the standard observation from the same night. In particular, we chose the red standard star HD19445 (spectral type: sdF) to mitigate second-order contamination. Telluric absorption was corrected using the template produced by normalizing the standard spectrum. More recently, we obtained three additional spectra using the Double Spectrograph (DBSP) mounted on the Hale 200-inch telescope at Palomar Observatory (P200) \citep{Oke1982}. For these observations, we used the dichroic D55, which splits the incoming light longer than or shorter than 5500~\AA\ into separate red and blue channels. We used a grism of 600 lines per mm blazed at 3780~\AA\ for the blue arm and a grism of 316 lines per mm blazed at 7150~\AA\ for the red arm. Both spectra were obtained using a slit width of $\rm 1.5\arcsec$. We reduce the observed P200/DBSP spectra using the Python Pypeit package \citep{Pypeit1, Pypeit2}, which can implement the standard reduction procedure of long-slit spectra highly automatically. We list the detailed information on the spectroscopic observations mentioned above in Table \ref{specobs} and plot them in Figure \ref{all_spec}. 

\begin{deluxetable*}{cccccccc}
\setlength{\tabcolsep}{0.06in}
\tablecaption{\it spectroscopic observations information \label{specobs}}
\tablewidth{0pt}
\tablehead{
DATE & Instrument & Grating  & slit-width & exposure time & S/N\tablenotemark{\footnotesize \rm 1} & resolution\tablenotemark{\footnotesize \rm 2}  & wavelength coverage  \\  
 &   &    &  arcsec    &   s   & $\rm pixel^{-1}$ & \AA & \AA}
\startdata
2018-01-15& FLWO/FAST &     300                      &  3    & 1800 &   24   & ---- & 3470-7400  \\
2018-02-11& FLWO/FAST &     300                      &  3    & 1500 &   32   & ---- & 3470-7400 \\
2022-06-29& P200/DBSP & 600/4000(blue),316/7500(red) &  1.5  & 1400 &   39   & 4    & 3100-10600\\
2022-09-02& P200/DBSP & 600/4000(blue),316/7500(red) &  1.5  & 600  &   63   & 4    & 3100-10600\\
2023-09-09& P200/DBSP & 600/4000(blue),316/7500(red) &  1.5  & 1600 &   66   & 4    & 3100-10600\\
\enddata
\tablecomments{ Information on spectroscopic observations taken for ASASSN-18ap. 
}
\tablenotetext{1}{The S/N was calculated around 6000 $\AA$ at observed wavelength.}
\tablenotetext{2}{This is the resolution for the blue arm of the P200/DBSP spectra, and it was evaluated from the FWHM of the emission lines in the lamp spectra }
\end{deluxetable*}

\begin{figure}[htb]
\figurenum{4}
\centering
\begin{minipage}{0.5\textwidth}
\centering{\includegraphics[angle=0,width=1.0\textwidth]{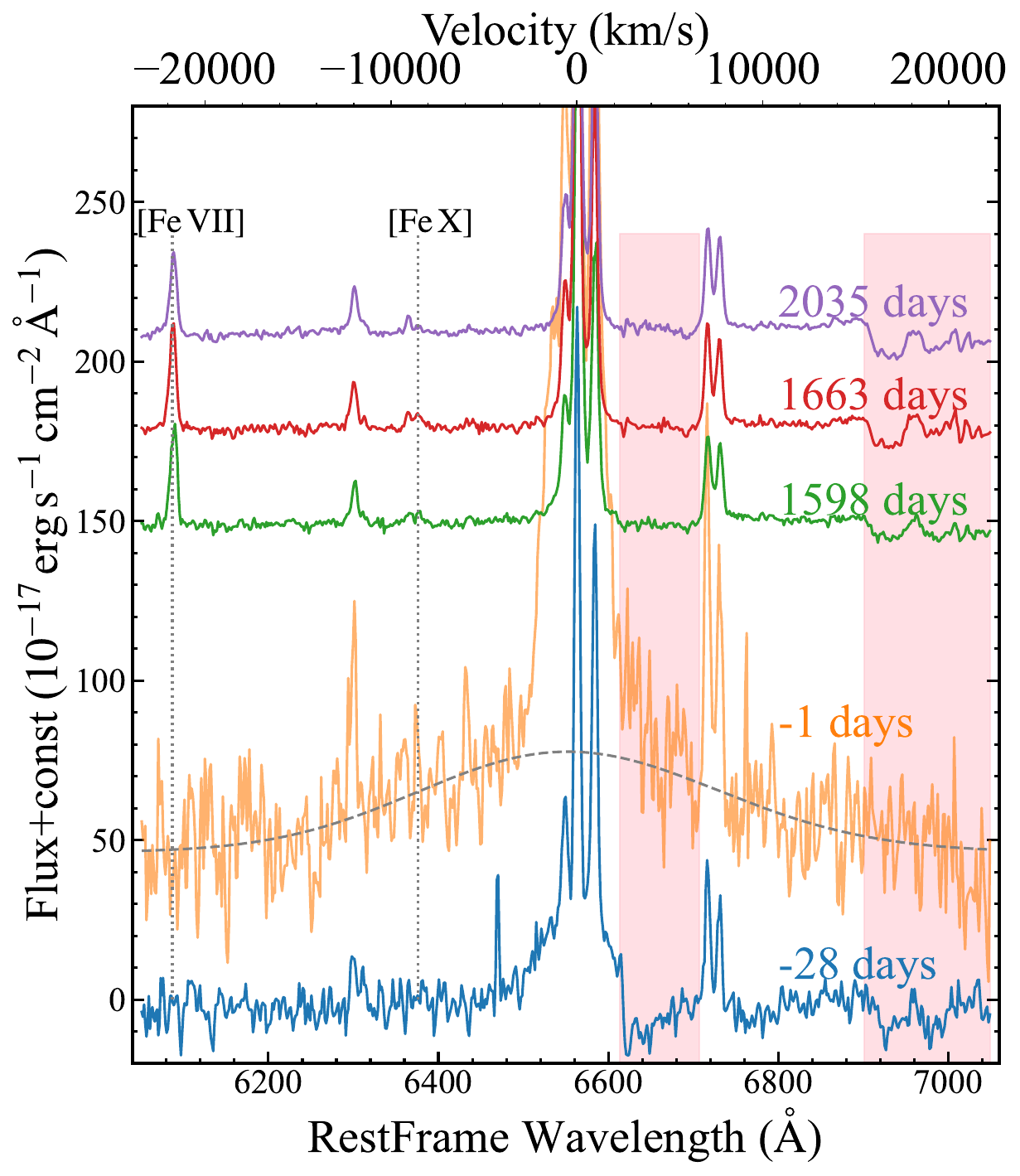}}
\end{minipage}
\caption{We presented the zoomed-in $\rm H\alpha$ region of the host-subtracted spectra. The phases relative to peak of each spectra were displayed on the right side using the same color scheme. Evidently, an intermediate width $\rm H\alpha$ component was observed in every spectra. The spectra taken around peak exhibited a very broad $\rm H\alpha$ component (indicated by the black dashed curve), and high-ionization lines denoted by the vertical dotted line emerged in the late-time spectra. \label{all_Ha}}
\end{figure}

\section{Data Analysis and Results} 
\label{sec3} 
\subsection{spectral data analysis} 
\label{sec3.1}
\subsubsection{The fitting process of continuum and emission lines}
\label{sec3.1.1}
To obtain detailed properties of the emission lines, we employed the following approach to model the continuum and emission lines. First, we corrected the spectra for the Galactic extinction using the dust map of \citet{dust_recalib} and the extinction curve of \citet{Fitzpatrick1999}. Subsequently, we utilized the TWFIT~\footnote{https://github.com/wybustc/twfit/} code to fit the continuum of ASASSN-18ap spectra. This code models the pseudo-continuum with a nonnegative linear combination of host stellar templates and an optional power law or black body component. To reduce the computational time of the fit, we used the six nonnegative independent components (ICs) compressed from the BC03 library of simple stellar populations (\citealt{Bruzual2003}) by \citealt{Lu2006}. Additionally, we incorporated a blackbody component to model the flux of the transient. We also took into account the potential dust attenuation effects of the host galaxy and the broadening of the stellar template caused by the stellar velocity dispersion. We masked the strong emission lines and the telluric regions during the fitting procedure. An example of the continuum fitting of a spectrum taken by P200/DBSP is shown in Figure \ref{spec_fit}.

After subtracting the continua from these spectra, the emission line spectra were modeled using a combination of Gaussian functions to measure the flux of narrow and broad lines. The fitting was performed using the python code MPFIT~\footnote{https://code.google.com/archive/p/astrolibpy/downloads}. Specifically, we divided each emission line spectrum into multiple line groups and fitted the lines in each group together. Each group either contain a single isolated emission line , or multiple lines that are blended together (e.g. $\rm H\alpha$ and $\rm [N\,\textsc{ii}]$ doublet) or have fixed theoretical flux ratio~\footnote{In our line fitting procedure, the flux ratio of  doublets $\rm [N\,\textsc{ii}]\lambda\lambda6583,6548$,  $\rm [O\,\textsc{iii}]\lambda\lambda5008,4959$,  $\rm [Ne\,\textsc{v}]\lambda\lambda3345,3425$,  $\rm [Ne\,\textsc{iii}]\lambda\lambda3869,3967$,  $\rm [O\,\textsc{i}]\lambda\lambda6301,6364$ at 2.96, 2.98, 0.37, 3.3, 3} (e.g., the $\rm [N\,\textsc{ii}]$ doublet). In general, we modeled each line in a group with a single narrow Gaussian function ($\rm FWHM<800\,km\,s^{-1}$) and ensured that the line widths and velocity shift relative to the theoretical wavelength for all the lines are the same in the fitting process within the same group. Additionally, a constant or linear function was sometimes added to the fitting process of each line group to account for possible residuals from the continuum fitting. Extra broad Gaussian functions ($\rm FWHM>1000\,km\,s^{-1}$) were only considered when modeling the profile of $\rm H\alpha$ and $\rm H\beta$, and a single broad Gaussian was sufficient to model the broad component detected in these two lines for all the spectra except the one taken around the peak light. For the spectrum taken around the peak, a second broad Gaussian was needed, as an additional broad wing was evident. We present examples of the emission line fitting in Figure \ref{line_fit}, and the derived emission line properties for the five spectra of ASASSN-18ap in Table \ref{spec_fit}. 
\subsubsection{spectral evolution}
\label{sec3.1.2}
\begin{figure*}[htb]
\figurenum{5}
\centering
\begin{minipage}{1.0\textwidth}
\centering{\includegraphics[angle=0,width=1.0\textwidth]{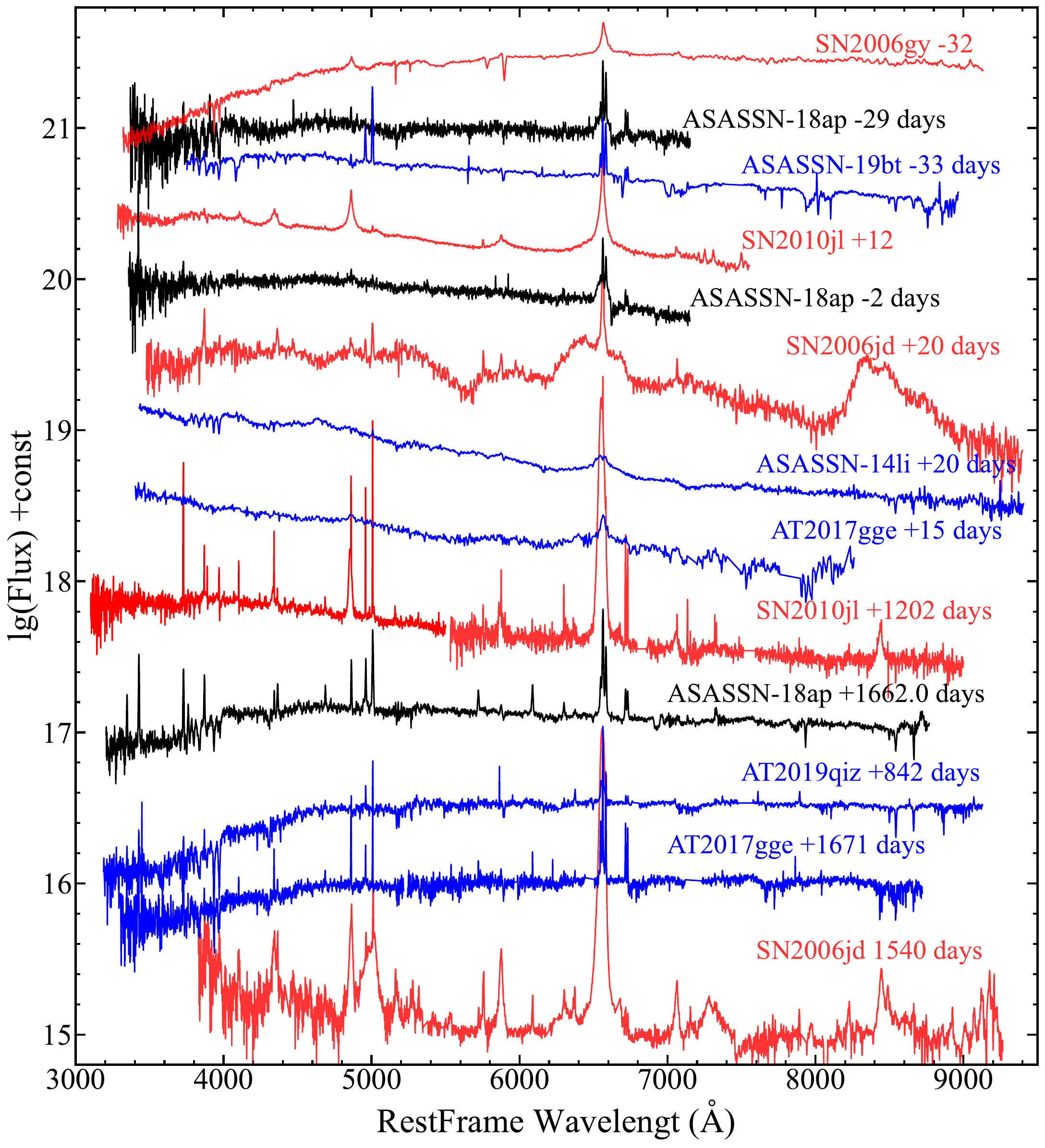}}
\end{minipage}
\caption{We compared the spectra of ASASSN-18ap at different phases (black curves) with the spectra of TDEs (blue curves) and SNe (red curves).The corresponding event and phase relative to the peak (SN2006jd relative to discovery, since the peak was not captured) are shown on the right side of the spectra with the same color. The gray dashed lines marked some low-ionization lines (e.g., $\rm O\,\textsc{i}$ and $\rm Ca\,\textsc{ii}$), typically observed in the late-time spectra of SNe. The green dashed lines marked several high ionization coronal lines.  The SNes used for comparison included  SN2006gy (e.g.\citealt{SN2006gy2007,SN2006gy}), SN2010jl (e.g., \citealt{SN2010jlCL1,SN2010jlCL2,SN2010jl_specdata}), SN2006jd (\citealt{SN2005ip&SN2006jd}). The TDEs used for the comparison included ASASSN-19bt (\citealt{ASASSN-19bt}), AT2019qiz (\citealt{AT2019qiz,AT2019qizCL}), AT2017gge (\citealt{Wang2022ApJL,Onori2022}, and ASASSN-14li 
(\citealt{Holoien2016}). Some of the data were also obtained from the Berkeley Supernova Ia Program (\citealt{supernovaeIaprogram}), WISeREP (\citealt{WISeREP}), and ESO science archive. \label{spec_comparison}}
\end{figure*}

The spectral analysis revealed that all five spectra had an intermediate-width broad $\rm H\alpha$ with the FWHM narrowing from an initial $\rm 4000\, km/s$ to $\rm 2000\, km/s$ at late-time (see Figures \ref{all_spec} and Figure \ref{all_Ha}). The $\rm H\beta$ showed this intermediate-width component only in the two early spectra. 
Additionally, the FLWO/FAST spectrum taken around the optical peak showed a blue continuum and a very broad component with $\rm FWHM\sim19000\,km/s$. We plotted the spectra of ASASSN-18ap at different phases in Figure \ref{spec_comparison}, comparing them with those from some typical TDEs and Type IIn supernovae. The trend of narrowing $\rm H\alpha$ FWHM is in agreement with the general pattern of evolution of the line width of TDEs that becomes increasingly narrow (e.g., \citealt{vV2020ssr,Charalampopoulos2022}), but more data points are needed to reach a conclusion. Furthermore, the presence of a very broad $\rm H\alpha$ component with $\rm FWHM>10000\,km/s$ was expected in the TDE spectra around the peak (e.g., \citealt{vV2020ssr}; see detailed comparison in \ref{sec4.2}). Spectroscopic features typically observed in normal SNe II, such as P-cygni features or low ionization metal lines (e.g., \citealt{Filippenko1997}), were not found in these spectra, and therefore the normal SNe-II scenario is not favored. However, it should be noted that the detected features of $\rm H\alpha$ mentioned above have also been observed in the spectra of Type IIn supernovae, with the intermediate-width component potentially arising from the CSM interaction and the emission of ejecta or electronic scattering contributing to the very broad component (see details in Section \ref{sec4.1.1}). Interestingly, the three late-time P200/DBSP spectra also featured strong high-ionization lines, such as $\rm [Fe\,\textsc{vii}]$ and $\rm He\,\textsc{ii}$, all of these lines being narrow and having an FWHM of a few times 100 km/s. Their luminosity ($\rm [Fe\,\textsc{vii}]\lambda6087\sim 10^{40}\,erg\,s^{-1}$) was consistent with those found in tidal disruption events, but was more luminous than those of supernovae (see Section \ref{sec4.1.1}). Moreover, the $\rm [O\,\textsc{iii}]$ doublet detected in late-time spectra brightened significantly compared to the two early spectrum (see Section \ref{sec4.3}). 

\subsubsection{Black Hole Mass Estimate}
\label{sec3.1.3}
We then try to assess the mass of the SMBH from the stellar velocity dispersion. Given that the procedure TWFIT primarily focused on the decomposition of various spectral components and the derived stellar velocity dispersion was approximate, we further evaluated the value using pPXF (\citealt{CE2004,Cappellari2017,Wevers2017}), which can use the full spectrum fitting to extract the stellar populations and kinematics through the Penalized PiXel-Fitting method, based on the results of TWFIT. First, we measured the instrument resolution as $\rm FWHM\sim 4 \AA$ for the blue-arm spectra of P200/DBSP ($\rm 1.5 \arcsec$ slit) using the lamp spectra. Then, we used the pPXF procedure and the included MILES spectral library (\citealt{Vazdekis2010}) to derive the velocity dispersion for the spectra taken on UT2022-09-02, which was one of the highest S/N spectra. Before the fitting, we corrected for Galactic extinction and subtracted the blackbody component obtained from the continuum fitting by TWFIT, as pPXF employs a fitting with pure starlight components. The fitting yields a velocity dispersion of $\rm \sigma_* \sim 76\pm17 \,km/s$ (see figure \ref{line_fit}), and thus a SMBH mass of $\rm 10^{6.65^{+0.38}_{-0.48}}\,M_\odot$~\footnote{For the error of black hole mass, we only considered the propagated error from velocity dispersion measurements and didn't include the intrinsic scatter of the $\rm M_{BH}-\sigma$ relationship used, which was estimated to be 0.29 dex in \cite{Kormendy&Ho2013} }, according to the Equation (7) in \cite{Kormendy&Ho2013}. This result is consistent with the BH mass values expected in TDEs (e.g., \citealt{Wevers2017,Wevers2019}).

\subsection{Host galaxy properties}
\label{sec3.2} 
\begin{figure}[htb]
\figurenum{6}
\centering
\begin{minipage}{0.5\textwidth}
\centering{\includegraphics[angle=0,width=1.0\textwidth]{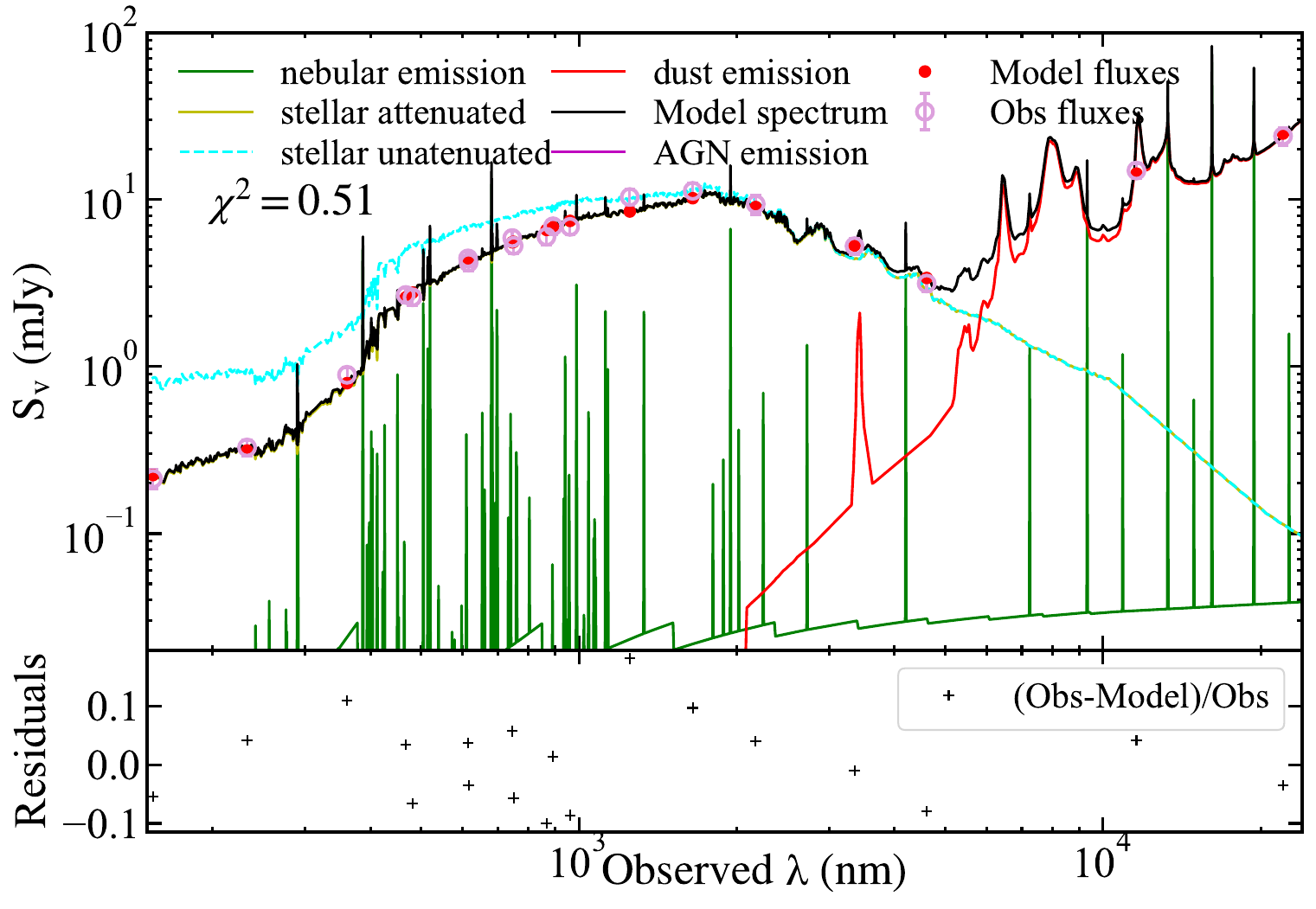}}
\end{minipage}
\caption{The host SED fitting results with the package CIGALE. In the top panel, we display the different components considered in the SED fitting (further details are provided in the main text), along with the best-fit SED achieved through their combination. The red points represent the model flux for each band derived from the best-fit SED, and the pink circle indicates the observed data. It is noteworthy that the AGN contribution to the best-fit SED is zero. In the bottom panel, we present the residuals between the observed data and model flux. \label{hostsed}}
\end{figure}

Reviewing the literature, there is no much available information about ASASSN-18ap's host galaxy except its morphology as a disk galaxy, and here we attempted to assess its activity status of AGN, stellar mass, and black hole mass through the SED fitting.
We collected multiband photometry for the host galaxy of ASASSN-18ap from several archives, including the NASA/IPAC Extragalactic Database (NED), the Sloan Digital Sky Survey (SDSS; \citealt{sdss}), and Pan-STARRS (\citealt{PS1}). Data are listed in the table in Appendix \ref{appendiceB}. To model the spectral energy distribution (SED) of the host galaxy, we use the Python package Code Investigating GALaxy Emission (CIGALE; \citealt{CIGALE}). CIGALE can fit an SED of a galaxy from FUV to radio and estimate its physical properties, such as star formation rate, through the analysis of likelihood distribution. In our fitting, we assumed a delayed star formation history (SFH) with an optional exponential burst and used the single stellar population of \cite{Bruzual2003}. We also take into account dust attenuation with a module based on \cite{Calzetti2000}. Additionally, dust emission is modeled using \cite{Dale2014}, and AGN emission is calculated with the model of \cite{Fritz2006}. In Figure \ref{hostsed}, we show the fit to all the photometric data we collected above, after correcting for the Galactic extinction. The SED can be well-fitted solely by stellar components without a significant contribution from the AGN component. This result is consistent with its $\rm W1-W2 \sim$ 0.11 (\citealt{Stern2012,Yan2013}), and non-variability before the outburst in the ATLAS-o band light curve. Actually, as we will discuss in section \ref{sec4.3}, the early spectrum of ASASSN-18ap placed it in the BPT diagram (\citealt{BPT,VO1987}) near the boundary between the star-forming galaxies region and the composite region, which also excluded the existence of significant AGN activity.

The mass of the SMBH in the galaxy center is estimated to be $\rm 10^{6.82}\,M_\odot$ based on the total stellar mass of $\rm \sim 10^{10.40}\,M_\odot$ obtained by the CIGALE SED fitting and its relation to the black hole mass (see Equation (4) in \citealt{Reines2015}). However, the mass is likely overestimated considering the disk dominated morphology of the host galaxy.

\subsection{Optical and infrared light curves}
\label{sec3.3} 
\begin{figure*}[htb]
\figurenum{7}
\centering
\begin{minipage}{0.9\textwidth}
\centering{\includegraphics[angle=0,width=1.0\textwidth]{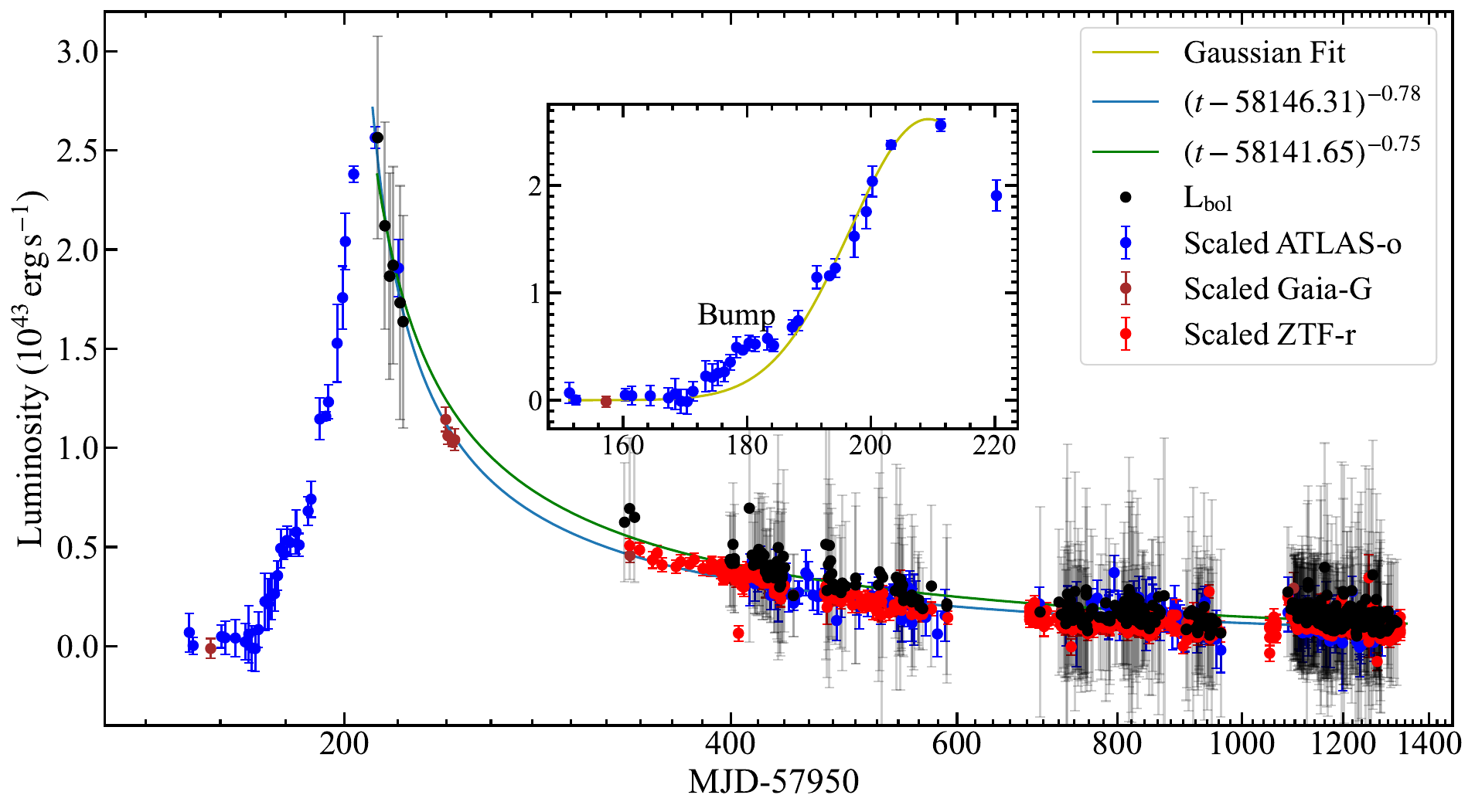}}
\end{minipage}
\caption{The bolometric light curve from the blackbody fitting to the multi-band photometry (black points). For comparison, we also show the monochromatic light curves: ATLAS-o (blue points), Gaia-G (brown points), and ZTF-r (red points). All of these monochromatic light curves were scaled to the Swift/UVOT blackbody luminosity in the overlap period. The green curve and blue curve represent the power-law fitting results for the decline phase of the bolometric light curve from the blackbody fitting results and from the one spliced by the different monochromatic light curves mentioned above, respectively. In the inserted panel, we zoomed in on the rising phase, which could be well fitted with a Gaussian function (yellow curve), and evidently there is a bump feature at the onset of the brightening. \label{wholeLC}}
\end{figure*}

We used a black-body model to analyze the physical parameters of the OUV radiation from ASASSN-18ap. We corrected the multi-bands light curves for Galactic extinction and subtracted the host contribution. We fit all the Swift six bands except the V-band, which was heavily diluted by starlight, in the vicinity of the peak. The outburst had a peak OUV luminosity of $\rm 10^{43.41}\,erg\,s^{-1}$ (i.e. the first Swift epoch). The black-body temperature remains relatively constant at approximately 10000 K for over a week around the peak, and the surface radius is on the order of $\rm 10^{15}\,cm$. For the rising phase, we simply scaled the monochromatic ATLAS-$o$ light curve to the Swift/UVOT blackbody luminosity because of the lack of multi-band photometry. As shown in Figure~\ref{wholeLC}, the rising phase roughly follows a Gaussian function, with an excessive bump that presents at the beginning of brightening (see more details in Section \ref{sec4.6}).

\begin{figure*}[htb]
\figurenum{8}
\centering
\begin{minipage}{1.0\textwidth}
\centering{\includegraphics[angle=0,width=1.0\textwidth]{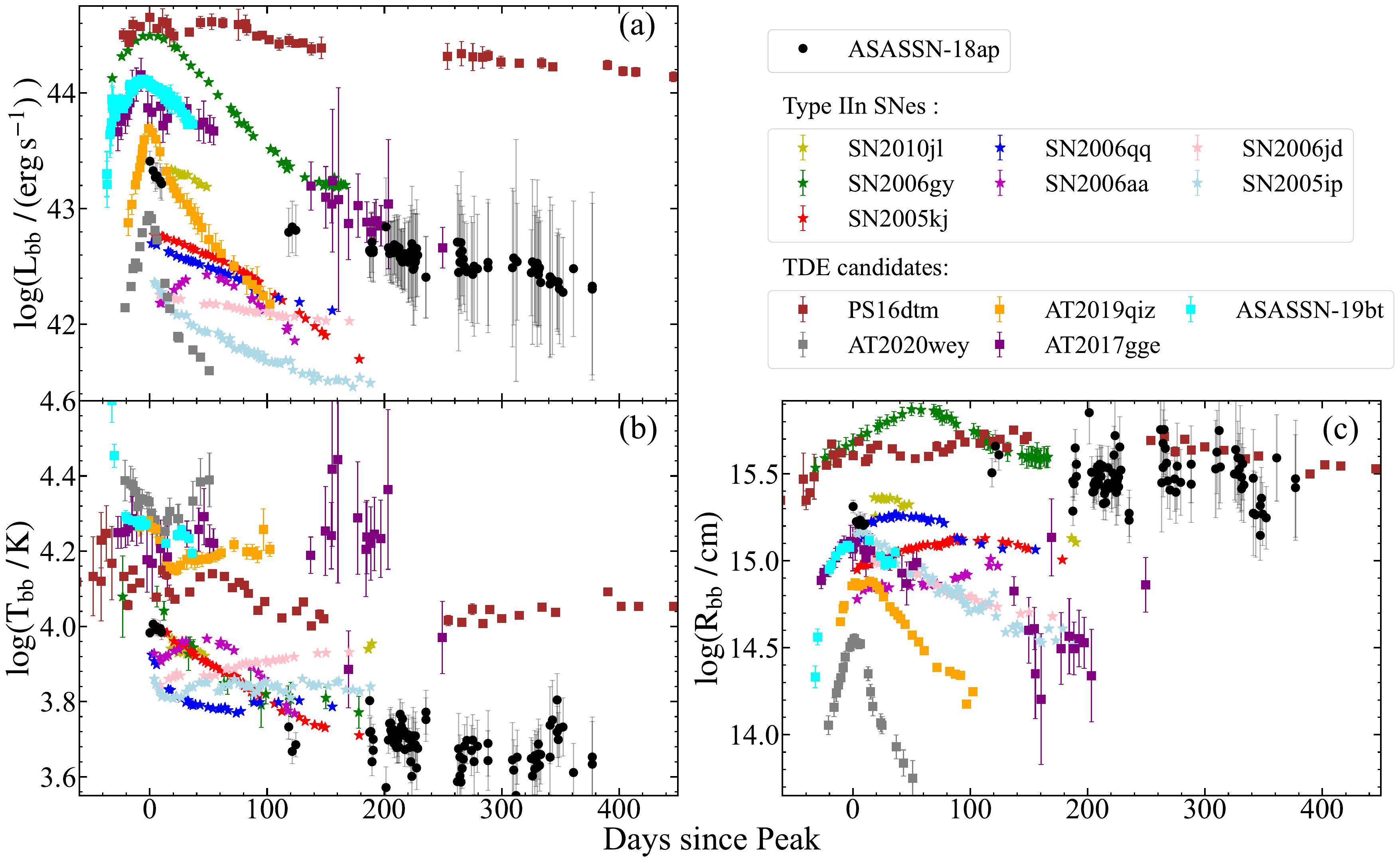}}
\end{minipage}
\caption{ Evolution of blackbody luminosity (panel a), temperature (panel b), and radius (panel c) for ASASSN-18ap (black points), in comparison with those from TDEs (squares) and SNes (stars). The black-body parameters of ASASSN-18ap at late times were modeled using only the available optical bands (see the main text). The SNe used for comparison include SN2010jl (e.g., \citealt{SN2010jlCL1, SN2010jlCL2}), SN 2006gy (e.g., \citealt{SN2006gy}), SN2006qq, SN2006jd, SN2006aa, SN2005ip, and SN2005kj. Data for the last four events were retrieved from \citealt{Taddia2013}, and we utilized the recalculated results of SN2010jl from \citealt{ASASSN-17jz}. The TDEs used for comparison include PS16dtm (\citealt{PS16dtmCL}) , AT2019qiz (\citealt{AT2019qiz}), ASASSN-19bt (\citealt{ASASSN-19bt}), AT2020wey (\citealt{AT2020wey}), AT2017gge (\citealt{Wang2022ApJL,Onori2022}) .   \label{LC_comparison}} 
\end{figure*}

We utilized two different methods to derive the bolometric luminosity and temperature evolution of the late time. First, we applied black-body fitting to ATLAS-o, ATLAS-c, ZTF-g, ZTF-r, and Gaia-G, with interpolation of all other photometry to the MJD grids of ZTF-g. Black-body parameters are shown in Figures \ref{wholeLC} and \ref{LC_comparison}. We also tested the situation where we binned the light curve into 20-day sections to increase the signal-to-noise ratio, and the result was the same. Swift photometry around the peak was used to calculate the blackbody luminosity, which roughly follows the power law formula of $\rm L_{bb}\propto(t-58141.65)^{-0.75}$ (Figure \ref{wholeLC}). Integrating the bolometric light curve yielded a total OUV radiation of $\rm 3.9\times10^{50}\,erg$. The blackbody temperature decreased rapidly from 10,000K at its peak to around 5000K, then remained constant at a later stage (phase$>$100 days). However, it is necessary to be cautious when considering this temperature evolution, as the five bands used in this study do not span a wide enough range of wavelengths to provide reliable black-body results. Additionally, the host's contribution may still be present in the photometry data, despite attempts to eliminate it. Furthermore, ZTF-r/ATLAS-o could be influenced by $\rm H\alpha$ emission. Therefore, the blackbody fitting with only these optical bands may result in a lower temperature than the actual one (see Appendix \ref{appendiceC} for more details). Secondly, we employed another method to estimate the evolution of luminosity, in which the bolometric light curve is generated by scaling the ATLAS-o, Gaia-G, and ZTF-r monochromatic light curves to the Swift/UVOT blackbody luminosity in the overlapped period. A power law with the formula $\rm L\propto(t-58146.31)^{-0.78}$ fits the decay of the light curve well, and the total OUV radiation energy was estimated as $\rm 3\times10^{50}\,erg$. Both of these results are similar to each other.

\begin{figure}[htb]
\figurenum{9}
\centering
\begin{minipage}{0.5\textwidth}
\centering{\includegraphics[angle=0,width=1.0\textwidth]{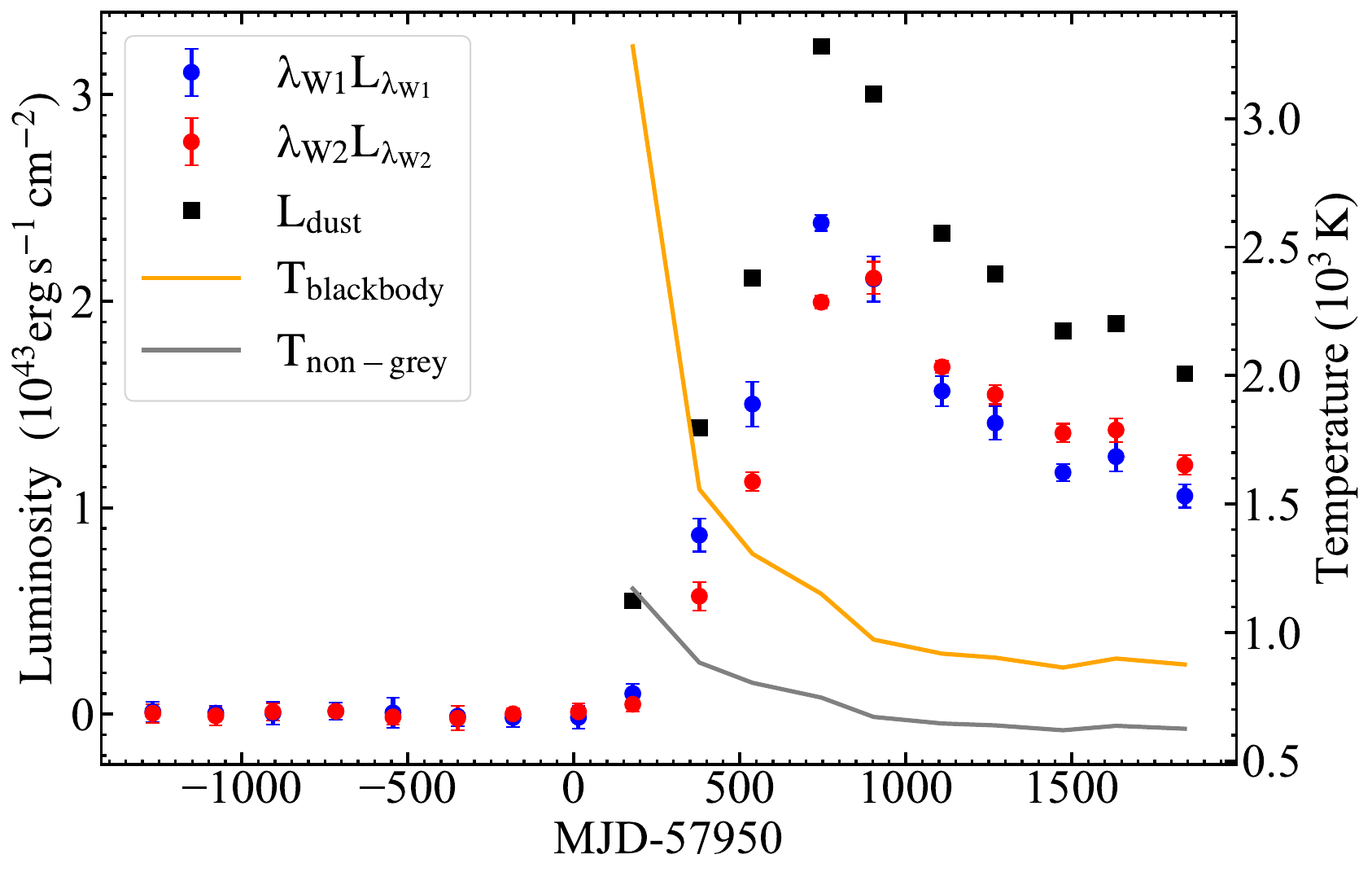}}
\end{minipage}
\caption{ The host-subtracted WISE light curve (blue points for W1 and red points for W2 in a formula of monochromatic luminosity) and the dust luminosity (black square) , which was evaluated by fitting a blackbody model to the W1 and W2 bands. The orange curve represents the temperature obtained from the blackbody model, while the gray curve shows the temperature considering the non-grey feature (see \ref{sec3.3}). Note that the temperature at the first brightening point could be contaminated by the simultaneous OUV radiation. The fitting results to the host-subtracted WISE flux are available as the data behind the figure.\label{WISE_fit}}
\end{figure}

We model the simultaneous host-subtracted W1 and W2 photometry of ASASSN-18ap with a single black body and present the results in Figure \ref{WISE_fit}. The infrared peak luminosity is about $\rm 10^{43.51}\,erg\,s^{-1}$, and the integrated infrared energy up to the latest WISE epoch is about $\rm 3.1\times10^{51}\,erg$. With black-body luminosity and temperature, the size of the emission region is estimated to be about 2$\times 10^{17}$ cm. The peak luminosity in infrared is higher than that in OUV, and the integrated infrared energy is about one order of magnitude higher than the optical one (hereafter referred to as infrared excess). In reprocessing models, the excess suggests that the intrinsic bolometric luminosity is greatly underestimated with the above OUV model, probably due to large dust extinction, an additional EUV bump in the SED, or anisotropic emission (see Section \ref{sec4.4}). Unfortunately, it is difficult to distinguish these possibilities solely on the basis of the available data. Future observations with better wavelength coverage are desired. To account for the non-gray nature of grains, we also try a modified blackbody of $\rm \lambda^{-\beta}\,B_\lambda(T)$, where the absorption coefficient varies with $\lambda$ in a powerlaw with index $\beta=1.7$ typical for the Galactic interstellar medium. In this case, the best-fit temperature is lower (see Figure \ref{WISE_fit}) while the overall IR luminosity is similar.


\subsection{Photoionization simulation for narrow line spectra}
\label{sec3.4} 
As shown in Figure \ref{all_spec}, high-ionization narrow emission lines emerged in late-time P200/DBSP spectra, and their intensities can be used to constrain the shape of the ionizing continuum and the physical properties of the line-emitting gas. It has been recognized that the typical SED of TDEs consists of a strong UV bump and a soft X-ray spectrum (black-body of a few $\rm \times 10\,eV$; \citealt{Saxton2020}), while the interaction between supernovae and the circumstellar medium will give rise to thermal plasma emission with temperatures of a few keV. Therefore, we performed simulations using the photoionization code CLOUDY v17.02 (\citealt{CLOUDY2017}) for two different input SEDs: (i) 4 SEDs calculated from the unified TDE model with different viewing angles (\citealt{Dai2018}), and we found that the smallest angle (5.7-22.5°) has the best result; (ii) an SED for Type IIn supernovae. The latter is represented by an optical bremsstrahlung tail joint with the plasma SED at a temperature of $\rm \sim 1\,keV$, which is calculated by the procedure PyatomDB (\citealt{2020Atoms...8...49F}). In our simulations, we assumed that the narrow lines are produced by clouds with similar physical conditions. We assumed that the clouds have a column density of $10^{23}$~cm$^{-2}$ and abundance of solar, and found that both have no significant influence with mild variation.  We varied hydrogen density $\rm n_H$ and ionization parameter $\rm U$ (i.e., $\rm \frac{\Phi_H}{ n_Hc}$, $\rm \Phi_H$ is ionizing flux) to make grids of simulated narrow line ratios ($\rm \eta_{sim}$) and compared them with observation by calculating $\rm \chi^{2}=\Sigma(\eta_{sim}-\eta_{obs})^2/\eta_{err}^2$, in which the $\eta_{obs}$ ($\rm \eta_{err}$) is the observed line ratios (ratio uncertainties~\footnote{The ratio uncertainties were propagated from the uncertainties of line intensities, which is a combination of the statistical error given by the emission line fitting procedure described in Section \ref{sec3.1.1} and a systematic error of 10\% of the line flux. The latter was estimated from the difference between the measured fluxes of the same line from two spectra taken at UT2022-09-02 and UT2022-06-29. }) derived from the spectrum taken at UT2022-09-02.

\begin{figure*}[htb]
\figurenum{10}
\centering
\begin{minipage}{0.9\textwidth}
\centering{\includegraphics[angle=0,width=1.0\textwidth]{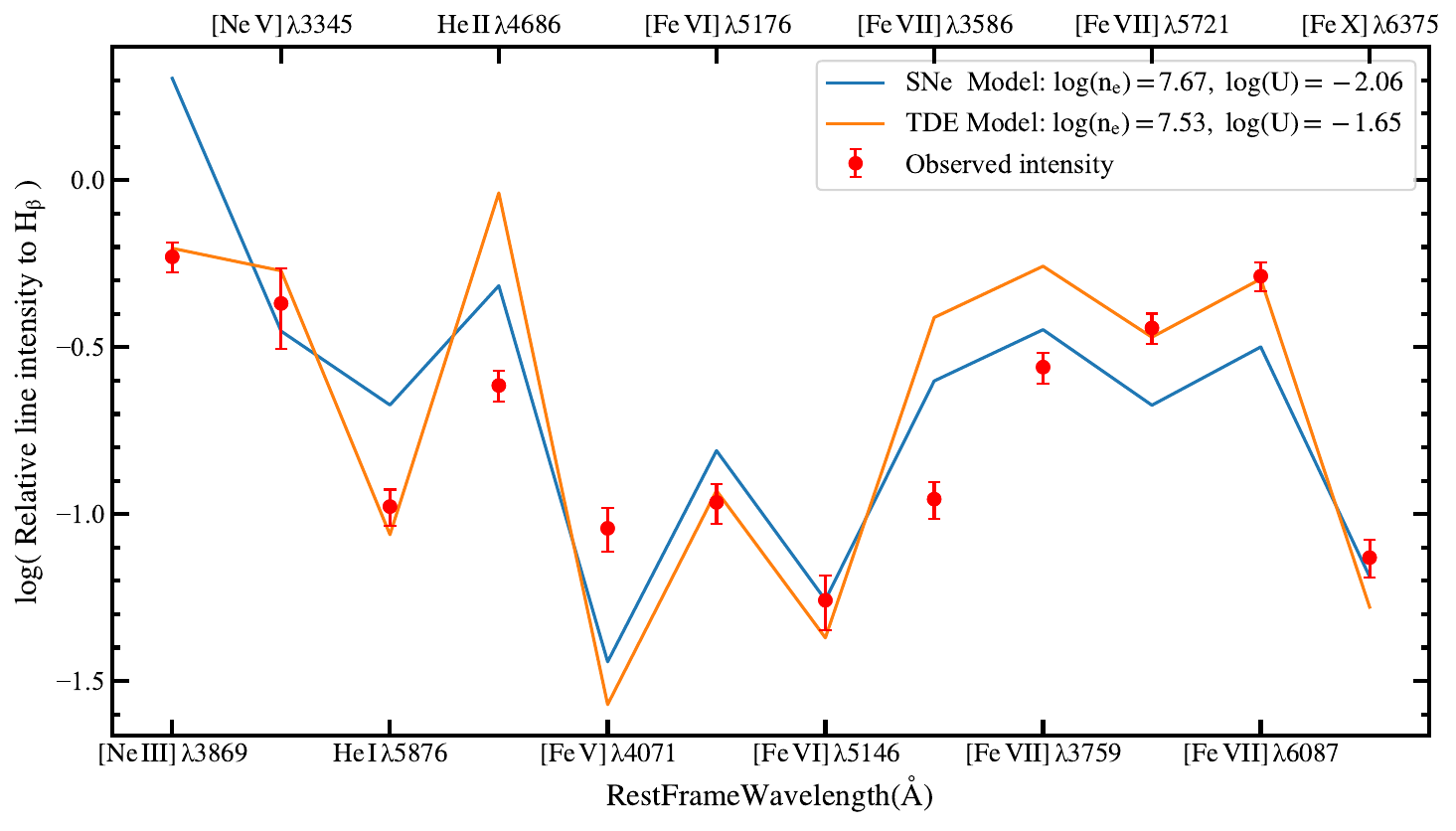}}
\end{minipage}
\caption{ We present the photoionization simulations for observed narrow line intensity with CLOUDY. Here, the line intensity (the red points) is normalized to that of $\rm H\beta$, and such a relative intensity was used as input for the CLOUDY simulation. The blue curve shows the result of the fitting of the SN scenario, while the orange curve represents the best TDE model with SED from \cite{Dai2018}. In addition, we list some key parameters derived from the simulation for the two models in the upper right corner of the figure. \label{cloudy}}
\end{figure*}

We compare the best-fitted line ratios produced by the simulations (by minimizing $\chi^2$) with those observed in Figure~\ref{cloudy}, where the density and ionization parameters of the models are also labeled. Both SEDs can qualitatively reproduce the observed narrow-line ratios, so our simulations cannot determine which of the two SEDs is more appropriate for ASASSN-18ap directly. The Monte Carlo method was used to calculate the emission area of the models, which was equivalent to a sphere with a radius of $8.4\pm0.79\times10^{17}$ cm (TDE) and $9.5\pm0.8\times10^{17}$ cm (SNe), by matching observed emission line fluxes. As described in Section~\ref{sec4.1.2}, the size is much larger than expected in a Type IIn supernova. Additionally, the SNe model requires an X-ray luminosity of $\rm L_{0.3-10\,KeV}=10^{43.44^{+0.07}_{-0.08} }$, which is at least one order of magnitude higher than that observed in Type IIn ($\rm \lesssim 10^{42}\,erg\,s^{-1}$, see \citealt{Ross2017,Chandra2018}), while the X-ray luminosity ($\rm L_{0.3-10\,KeV}= 10^{42.56^{+0.08}_{-0.09}}$) demanded by the TDE model was consistent with observations (e.g. \citealt{Saxton2020, Gezari2021, Guolo2023}). 

On the other hand, the gas density and temperature can be constrained by the ratios of narrow lines of $\rm [Fe\,\textsc{vii}]$ or $\rm [O\,\textsc{iii}]$ directly~\citep{Osterbrock2006}. We computed the line ratios for a grid of temperatures ($\rm 10^{3}-10^{5}\, K$) and densities ($\rm 10^{1}-10^{10}$ cm$^{-3}$) using the IDL package CHIANTI v10.1.3 (\citealt{Chianti1997,Chianti}). Then, a pair of observed line ratios, $\rm \frac{[Fe\,\textsc{vii}]\lambda3759}{[Fe\,\textsc{vii}]\lambda6087}$ and $\rm \frac{[Fe\,{vii}]\lambda5160}{[Fe\,{vii}]\lambda6087}$, would reveal the density and temperature that produce exactly these two ratios by minimizing $\chi^2$. We estimated the errors using a Monte Carlo method, perturbing the observed line ratios 10,000 times according to the measurement error. This gave us a logarithmic density of $\rho \rm (cm^{-3}) \sim 6.12^{+0.48}_{-0.42}$ (with an error of 90\%) and a temperature of $\rm \sim 10^{4}\,K$, which is roughly consistent with the results of the CLOUDY simulations. In addition, we studied the pair $\rm \frac{[Fe\,\textsc{vii}]\lambda3759}{[Fe\,\textsc{vii}]\lambda6087}$ and $\rm \frac{[O\,\textsc{iii}]\lambda4363}{[O\,\textsc{iii}]\lambda5007}$, which yielded a logarithmic density of $\rho \rm (cm^{-3})\sim 7.20^{+0.39}_{-0.50}$ and a similar temperature of $\rm \sim 10^{4}\,K$.

\subsection{Fitting multi-wavelength light curves with MOSFiT} 
\label{sec3.5} 

We attempted to fit the multiband host-subtracted light curves using the Modular Open-Source Fitter for Transients (MOSFiT; \cite{Guillochon2018}). This tool generates Monte Carlo ensembles of semi-analytical light-curve fits to the data sets and returns their associated Bayesian parameter posteriors. Before fitting, we binned the ZTF-r and ATLAS-o light curves at a late time to reduce the number of data points. The MOSFiT fit was only applied to the data before $\rm MJD\,58600$. We considered two different built-in models: CSMNI, and TDE. The CSMNI model is powered by the combination of the CSM interaction ( \citealt{Chatzopoulos2013,Villar2017,JiangCSM}) and nickel-cobalt decay (\citealt{Nidecay}). The TDE model in MOSFiT was described by \citealt{Mockler2019}, which converts the fallback rate of the material onto the black hole post-disruption directly to bolometric luminosities with constant efficiency and passes these luminosities through the viscosity and reprocessing transformation function to create multiwavelength light curves.

\begin{figure*}[htb]
\figurenum{11}
\centering
\begin{minipage}{1.0\textwidth}
\centering{\includegraphics[angle=0,width=1.0\textwidth]{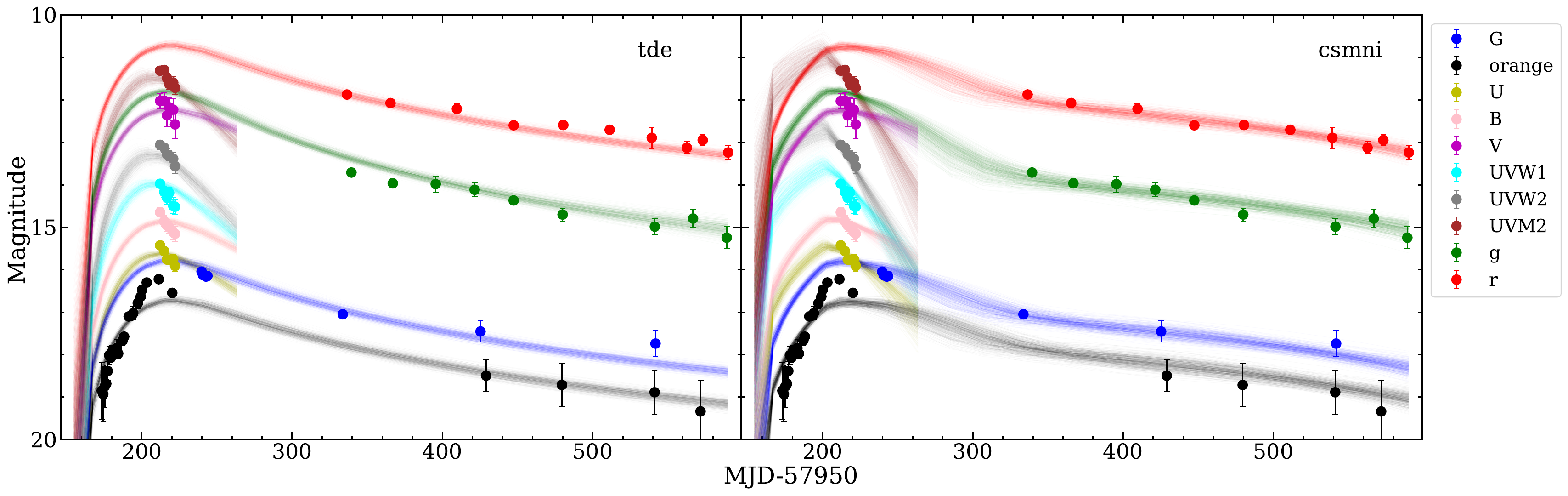}}
\end{minipage}
\caption{The MOSFiT fitting to the multi-band host-subtracted light curves. The left, and right panels show the fitting results of the TDE, and CSMNI models, respectively. The color coding used in the two panels is identical. The light curves constructed by MOSFiT are coded in the same way as the data points.  \label{mosfit_all}}
\end{figure*}

The priors of the model parameters are set as log-uniform or uniform and are summarized in Table \ref{MOSFiT}. Specifically, the priors of the CSMNI model were chosen to be similar to those in \cite{Kangas2022} and for the TDE model, we set the priors based on \cite{Mockler2019}. It's worth noting that we varied the slope of CSM density profile between 0 and 2 in the CSMNI model. We ran the two models until convergence using dynamic nested sampling with DYNESTY (\citealt{Speagle2020}). We show the fitting results of the light curve in Figure \ref{mosfit_all}, and the posterior distribution of the model parameters in Table~\ref{MOSFiT} and the corner plots~\ref{Cornerplot}. Both the two models provided a rough overall fit to the multiband light curves, but exhibited poor performance around the peak. Furthermore, the CSMNI model required an unreasonably high ejecta mass of $\rm \sim 68\,M_\odot$ (see Table \ref{MOSFiT} ). For the TDE model, the MOSFiT yielded a black hole mass $\rm log\,M_{BH}\sim6.77^{+0.12}_{-0.11}$~\footnote{The uncertainty here only consider the statistic error without the systematic uncertainty. \cite{Mockler2019} estimated a systematic error $\rm \sim 0.2\,dex$ for the black hole mass. }, similar to the values derived in Section \ref{sec3.1.3} using the $\rm M-\sigma$ relation or the relation with the total mass of the galaxy. The scaled impact parameter $\rm b\sim~0.99^{+0.08}_{-0.09}$ implies that the star ($\rm M_*\sim 1.03^{+0.59}_{-0.42} M_\odot$) was nearly fully disrupted by the black hole. In particular, the TDE model fitting yielded a high host galaxy hydrogen density of $\rm nH=20.66^{+0.26}_{-0.86}\,cm^{-2}$, corresponding to a significant host extinction of $\rm A_{v}=0.25^{+0.21}_{-0.22}$ (\citealt{PS1995,Guillochon2018}). 


\section{Discussion} 
\label{sec4} 
Both the SED fitting by CIGALE and the WISE W1-W2 color of ASASSN-18ap have eliminated a significant AGN contribution to the host galaxy (see Section \ref{sec3.1}), making it most likely that the outburst of ASASSN-18ap was caused by either a supernova or a tidal disruption event. We discussed these two scenarios in detail below, and the possibility of AGN variability was also further investigated.

\subsection{SN Scenario}  
\label{sec4.1} 

ASASSN-18ap was reported as a type II SN by \cite{ASASSN18ap-class} based on the presence of a broad H$\alpha$ in the optical spectrum taken immediately after its discovery.
However, it does not exhibit either P-Cygni lines in its early spectrum or low-ionization metal lines in the late-time spectra, which are typical for a normal SNe II (e.g., \citealt{Filippenko1997}). Additionally, the timescale of normal SNe II, which is usually months to a year, was significantly shorter than that of ASASSN-18ap. Therefore, normal SNe II can be largely excluded for ASASSN-18ap.  There is a type of supernova, known as Type IIn (e.g., \citealt{Smith2017}), that explodes in a dense circumstellar medium (CSM). In these events, the interaction of the ejecta with the circumstellar medium would modify the SNe's spectroscopic and photometric properties, thus bearing characteristics similar to those of ASASSN-18ap, as discussed in the following.

We compared the spectral features of ASASSN-18ap with those of typical Type IIn supernovae SN 2006gy (\citealt{SN2006gy}), SN2010jl (\citealt{SN2010jlCL1,SN2010jlCL2}), and SN2006jd (\citealt{SN2005ip&SN2006jd}) in different phases, as shown in Figure \ref{spec_comparison}. The figure reveals that Type IIn supernovae can display strong intermediate-width broad Balmer lines ($\rm few\ 10^{3}\,km\,s^{-1}$) similar to those of ASASSN-18ap in the pre-peak, peak, and post-peak phases. These intermediate width emissions are usually thought to be generated in the cold dense shell between the forward and reverse shock (e.g., \citealt{Chugai2004,Smith2017}). At the peak, a very broad ($\rm FWHM\sim19000\,km\,s^{-1}$) $\rm H\alpha$ component was detected in ASASSN-18ap, which is also commonly seen in Type IIn supernovae and can be attributed to the electronic scattering broadening effect from photon-ionized CSM and emission from high-velocity ejecta (e.g., \citealt{Smith2017}). However, the latter can be completely hidden as a result of the high opacity of the CSM, especially at an early stage.
Narrow emission lines from the circumstellar medium are also a typical feature of Type IIn supernovae (e.g.\citealt{Filippenko1997}), and in ASASSN-18ap, they may be heavily contaminated by emission lines from the host galaxy. At late times, in ASASSN-18ap we clearly see the emergence of high coronal lines (e.g., iron coronal lines $\rm [Fe\,\textsc{vii}]$), which are also detected in some Type IIn supernova with the CSM interaction providing high-energy ionization photons, such as SN2005ip (\citealt{SN2005ip}), SN2006jd (\citealt{SN2005ip&SN2006jd}), and SN2010jl (\citealt{SN2010jlCL1,SN2010jlCL2}). However, as we will discuss in Section \ref{sec4.1.1}, the coronal lines detected in ASASSN-18ap were at least one order of magnitude higher than those observed in Type IIn supernovae. Another discrepancy, as shown in Figure \ref{spec_comparison}, would be the absence of low-ionization metal lines for ASASSN-18ap, such as $\rm Ca\,\textsc{ii}$ Triplet and $\rm O\,\textsc{i}$, particularly at late time (e.g., 1542 days after explosion for SN2006jd, a phase similar to our late-time spectra for ASASSN-18ap), at which the density of CSM drops significantly, and thus the CSM becomes more transparent to the inner normal SNe emission. However, this discrepancy can be mitigated for ASASSN-18ap by a persistent strong CSM interaction at late times, or these lines may be too faint at such late times to be detected. Furthermore, some sources, such as SN2003ma (\citealt{Rest2011}) and SN2012ab (\citealt{SN2012ab_2018, SN2012ab_2020}), do not show such lines.

The evolution of Type IIn light curves shows a wide range of behaviors influenced by the mass-loss history of the progenitor (\citealt{Kiewe2012,Taddia2013}). Unlike normal supernovae, Type IIn supernovae can exhibit light curves as luminous and long-lasting as ASASSN-18ap, such as SN2006gy (\citealt{SN2006gy} and SN2010jl (\citealt{SN2010jlCL1, SN2010jlCL2}), since SNe IIn are also powered by the conversion of kinetic energy of the ejecta to radiation through CSM interaction in addition to radioactive decay (e.g., \citealt{Chugai1990,Smith2017,Chandra2018}). Figure \ref{LC_comparison} demonstrates that Type IIn supernovae can display blackbody luminosity, temperature, and radius similar to ASASSN-18ap in both aspects of magnitude and evolution behavior. In particular, post-peak temperatures for Type IIn supernovae are typically $\rm \lesssim 10000\,K$ with diverse evolution behavior such as increasing, decreasing or remaining roughly constant (e.g. \citealt{Taddia2013}). The blackbody temperature of ASASSN-18ap closely resembles that of Type IIn, although the late-time temperature of ASASSN-18ap may be underestimated.  Additionally, long-lasting infrared echoes have also been observed for some Type IIn supernovae (e.g. \citealt{Szalai2019}) such as SN2010jl (e.g. \citealt{Fransson2014}), with timescales and dust temperatures similar to those of ASASSN-18ap. However, the infrared peak luminosity of ASASSN-18ap was at least one magnitude higher than any supernova previously observed, as discussed in Section~\ref{sec4.1.2}. 


\subsubsection{The development of highly luminous coronal lines}
\label{sec4.1.1} 
\begin{figure}[htb]
\figurenum{12}
\begin{minipage}{0.5\textwidth}
\centering{\includegraphics[angle=0,width=1.0\textwidth]{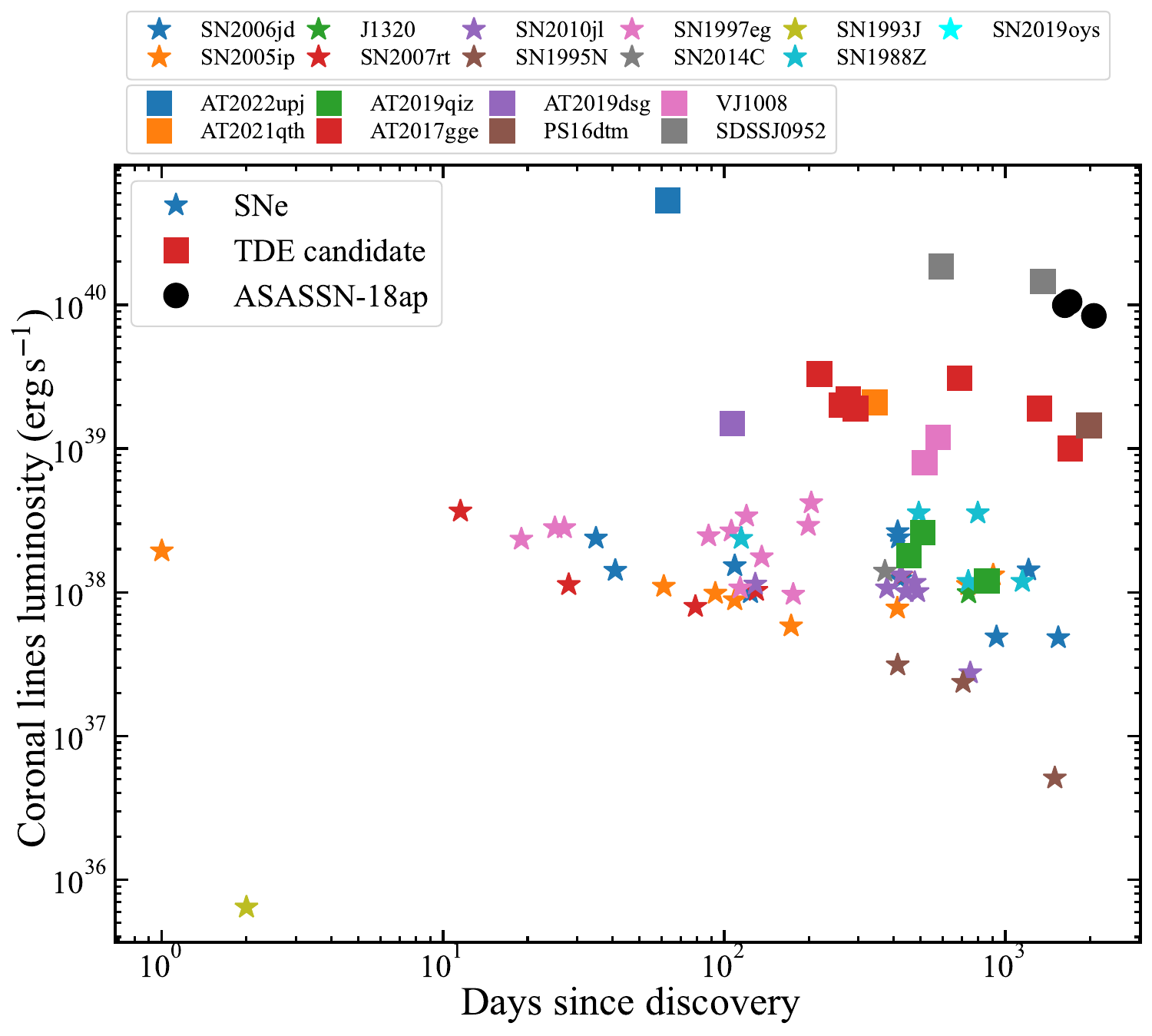}}
\end{minipage}
\caption{We compared the luminosity of ASASSN-18ap's coronal line (black point) with those detected in the SNes (stars) and TDE candidates (squares). The luminosity of the coronal line is calculated from $\rm [Fe\,\textsc{vii}]\lambda6087$ or $\rm [Fe\,\textsc{X}]6374$. The SNe included in the comparison are SN2005ip (\citealt{SN2005ip}), SN2006jd (\citealt{SN2005ip&SN2006jd}), SN2007rt (\citealt{SN2007rtCL}), J1320 (\citealt{J1320CL}), SN2010jl (\citealt{SN2010jlCL2,SN2010jlCL1}), SN1995N (\citealt{SN1995NCL}), SN1997eg (\citealt{SN1997egCL}), SN2014C (\citealt{SN2014CCL}), SN1993J (\citealt{SN1993JCL}), SN1988Z (\citealt{SN1988ZCL}), SN2019oys (\citealt{SN2019oysCL}). The TDE candidates included are AT2022upj (\citealt{AT2022upjCL}), AT2017gge (\citealt{Wang2022ApJL,Onori2022}), AT2019dsg (\citealt{AT2019dsgCL}), AT2019qiz (\citealt{AT2019qizCL}), PS16dtm (\citealt{PS16dtmCL}), VJ1008 (\citealt{VJ1008VJ2012CL}), SDSSJ0952 (\cite{SDSSJ0952CL1,SDSSJ0952CL2}) and AT2021qth (\citealt{AT2021qthCL}). For the above SNe or TDE candidates, the line luminosity of AT2022upj, AT2021qth, AT2017gge, PS16dtm, AT2019dsg, SN1997eg, SN2010jl, SN2007rt are calculated by this work with the spectra data retrieved from Open Astronomy Catalog API (https://github.com/astrocatalogs/OACAPI), TNS, and ESO Science Archive Facility (http://archive.eso.org/cms.html).\label{coronallines}}
\end{figure}

The spectra of ASASSN-18ap feature the late emergence of narrow high-ionization emission lines, particularly iron coronal lines such as $\rm [Fe\,\textsc{vii}]\lambda6087$ and $\rm [Fe\,\textsc{x}]\lambda6736$. These coronal lines have been observed in some SNe IIn, such as SN2005ip and SN2006jd, with the CSM interaction providing ionization photons. To compare the coronal lines of ASASSN-18ap with those in Type IIn supernovae, we have collected known supernovae with detected coronal lines from the literature and plotted their luminosities in Figure \ref{coronallines}. It is evident that the most luminous iron coronal lines detected in SNe are only a few times $\rm 10^{38} \,erg\,s^{-1}$, which is more than one order of magnitude lower than that of ASASSN-18ap (about $\rm 10^{40}\,erg\,s^{-1}$). In fact, as discussed in section \ref{sec3.4}, producing coronal lines with such high luminosity requires an X-ray luminosity of $\rm L_{0.3-10\,KeV}=10^{43.44^{+0.07}_{-0.08} }$, which is at least one order of magnitude higher than that observed in Type IIn ($\rm \lesssim 10^{42}\,erg\,s^{-1}$, see \citealt{Ross2017,Chandra2018}). However, coronal lines with luminosity comparable to ASASSN-18ap have been found in previous TDE candidates (see Figures \ref{coronallines} and \ref{sec4.2}).

Photoionization simulations of the narrow line spectrum of ASASSN-18ap, on the other hand, yielded an emission region much larger than the scale of ejecta as estimated below. For the type IIn SNe, the width of very broad Balmer lines is attributed to the Doppler broadening of the expansion of the ejecta, or the broadening of electron scattering in the ionized CSM,  or a combination of both (e.g., \citealt{Smith2017}). Hence, we assumed the ejecta expansion velocity at early phase ($\rm v_{s0}$) to be $\rm19000\,km/s$, corresponding to the FWHM of the very broad component detected in the peak spectra. Although the profile of this broad wing may be dominated by the electron scattering, an ejecta velocity of $\rm 19000\,km/s$ is a sufficiently high value for Type IIn (\citealt{Taddia2013}), allowing for a conservative estimation as follows. At $\sim1700$ days after the eruption, the velocity decreased to about $\rm v_{s1} \sim 2000\, km\,s^{-1}$, estimated from the FWHM of the intermediate width $\rm H\alpha$ in the spectra taken at UT2022-09-02 (i.e., the one used for CLOUDY simulations), where scattering wings are not found as expected at such a late time. In fact, the ejecta would decelerate quickly to a few times $\rm 1000\,km\,s^{-1}$ upon encountering the dense circumstellar medium (CSM), converting its kinematic energy into thermal energy. Then, at $\rm \sim1700$ days, the scale of ejecta, representing the emitting radius of the narrow lines, is estimated to be $\rm R\lesssim(v_{s0}+v_{s1})/2*t\sim1.54\times10^{17}\,cm$, which is approximately one order of magnitude lower than the value ($9.5\pm0.8\times10^{17}$) obtained from the photoionization simulation with the assumption of a covering factor $\rm \sim 1$.

\subsubsection{A high infrared luminosity }
\label{sec4.1.2} 
\begin{figure}[htb]
\figurenum{13}
\centering
\begin{minipage}{0.5\textwidth}
\centering{\includegraphics[angle=0,width=1.0\textwidth]{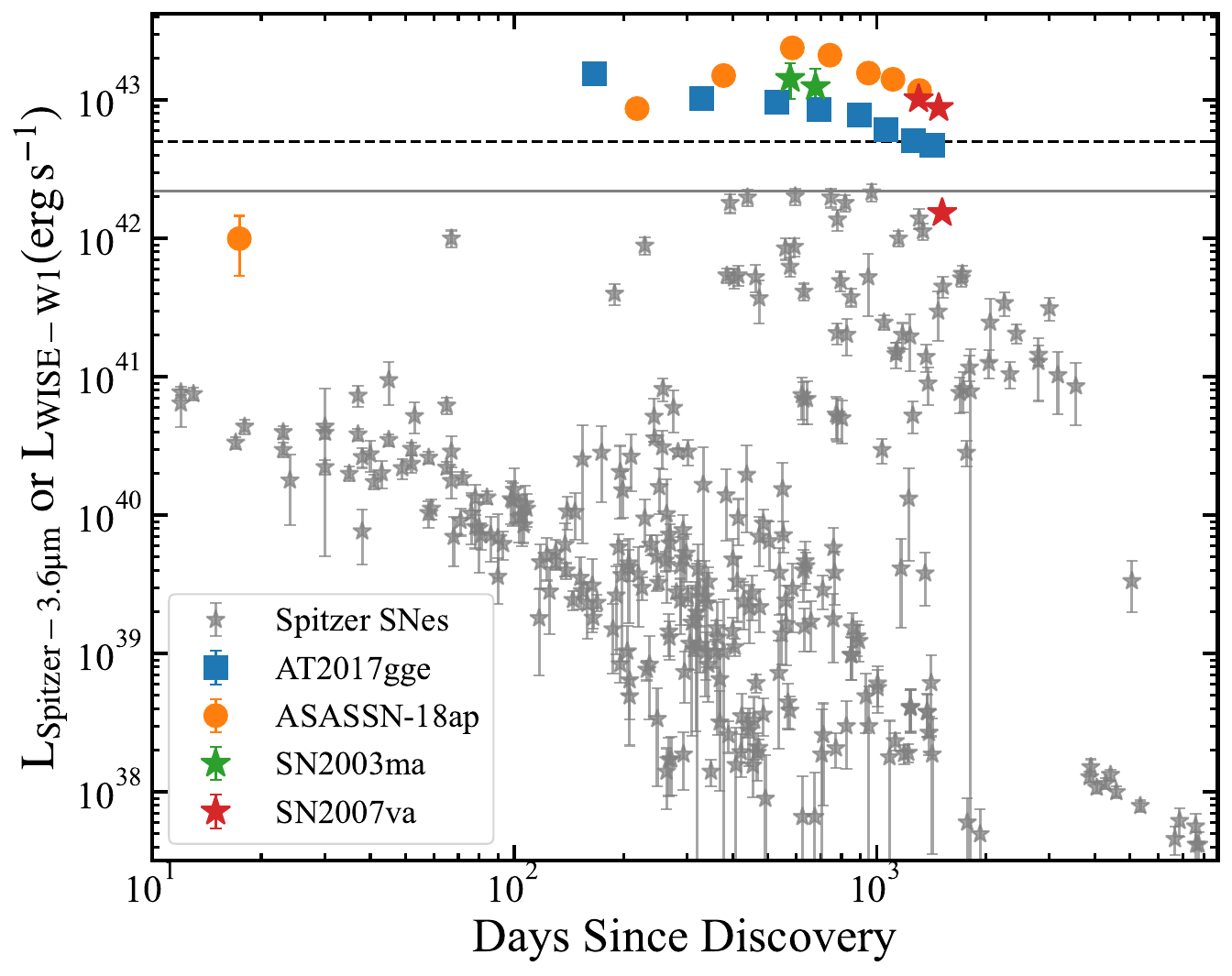}}
\end{minipage}
\caption{ 
The WISE-W1 light curve of ASASSN-18ap (orange points) compared to the 3.6$\rm \mu m$ light curve of Spitzer SNes (gray stars) from \cite{Szalai2019}. The upper boundary of the monochromatic infrared luminosity of the Spitzer SNes is marked with a gray horizontal line. Additionally, the maximum infrared luminosity of supernovae reported by \cite{SNeI-IR} is shown with a black dashed horizontal line. For comparison, we also included the infrared light curve of a dusty TDE AT2017gge (\citealt{Wang2022ApJL}), as well as two known SNes, SN2007va (\citealt{SN2007va}) and SN2003ma (\citealt{Rest2011}) that both have extraordinarily high infrared luminosity.  \label{IRcomparison} }
\end{figure}

A high infrared luminosity, i.e., $L>10^{42.5}$~\lum, has been proposed as a potential independent criterion to rule out the SNe scenario for IR-selected nuclear transients (e.g., \citealt{Wang2018,Jiang2019}). Recently,
\cite{SNeI-IR} studied the infrared luminosity of the Type I, Type II (including Type IIn) and SNe candidates using the WISE database and found that most SNe had a luminosity in the range of $\rm 10^{40.3} <$$\rm \lambda L_{W1}$~\footnote{\cite{SNeI-IR} acquired an absolute magnitude range $\rm -23<absW1<-17$, and here for comparison we converted it to monochromatic luminosity.} $\rm < 10^{42.7}\,erg\,s^{-1}$ (see Figure \ref{IRcomparison}). ASASSN-18ap and AT2017gge were identified as potential TDE candidates because of their much higher IR luminosity than other supernovae. \cite{Szalai2019} found similar limits $\rm \lambda\,L_{4.5\,\mu m}\lesssim10^{42.4}\,erg\,s^{-1}$ when examining SNe in Spizter images (see Figure \ref{IRcomparison}). In addition, \cite{Sun2022} studied the mid-IR emission of 10 SLSNe detected by WISE and found that the most luminous one had a luminosity $\rm \lambda\,L_\lambda \leq \rm 10^{42}\,erg\,s^{-1}$ in their sample. 

In addition to investigating the mid-infrared detection of known SNe, several projects (e.g., \citealt{Fox2021,Jencson2019,Kasliwal2017}) were designed to search for dust-obscured SNe. These projects have not yet discovered any SN with infrared luminosity as high as ASASSN-18ap. As far as we are aware, only two supernovae, SN2003ma (\citealt{Rest2011}) and SN2007va (\citealt{SN2007va}), have infrared luminosities similar to ASASSN-18ap, with values of $\rm 1\times 10^{43}\,erg\,s^{-1}$ and $\rm 6\times10^{43}\,erg\,s^{-1}$, respectively (see Figure \ref{IRcomparison}). In the first instance, the influence of the host galaxy has not been taken into account, whereas for the second, the possibility of a TDE cannot be ruled out. The authors of \cite{SN2007va} rejected the TDE hypothesis for SN2007va because the temperature of the mid-infrared radiation was much lower than what would be expected from the photospheric envelope. Nevertheless, this comparison is not valid as the infrared emission is more likely to be the result of the delayed dust reprocessing of the intrinsic emission, rather than from TDE radiation itself. Furthermore, recent studies of the IR echoes of TDEs (e.g., \citealt{Jiang2016, vV2016, Jiang2021ApJ, Wang2022ApJL}) indicate that the mid-infrared emission of SN2007va conforms well to that of TDEs, and thus the TDE scenario cannot be excluded. Therefore, the monochromatic peak luminosity of ASASSN-18ap in the infrared is approximately ten times brighter than the most luminous supernovae observed in the infrared (see Figure \ref{IRcomparison}).

\subsection{The TDE scenario} 
\label{sec4.2} 
In this section, we compare the observed photometric and spectroscopic properties of ASASSN-18ap with those of previously identified TDE candidates.

The optical light curve of ASASSN-18ap rises to a peak luminosity of $\rm \sim 10^{43.41} erg/s$ in about 40 days and then decays according to a power-law with an index of $\sim -0.75$ (see Figure \ref{wholeLC}). As depicted in Figure \ref{LC_comparison}, both the temporal evolution behavior and the luminosity range are consistent with the optical TDE scenario (e.g.,\citealt{vV2021ApJ,Gezari2021,Hammerstein2023}). The IR light curve peaks at a luminosity of about $\rm 10^{43.51}\,erg\,s^{-1}$ with a delay of 500 days to the optical peak, which agrees well with the properties of IR echoes from dusty TDEs (e.g., AT2017gge; \citealt{Onori2022,Wang2022ApJL}). The total OUV energy observed for ASASSN-18ap is approximately $\rm 2\times10^{50}\,erg$, which is at the lower end of known TDEs and very similar to some sources, such as iPTF16axa ($\rm 5.5\times10^{50}\,erg$, \citealt{Hung2017}), ASASSN-14ae ($\rm 1.7\times10^{50}\,erg$, \citealt{Holoien2014}). 
What distinguishes ASASSN-18ap is its lower blackbody temperature of OUV radiation and a possible declining trend in temperature (see Section \ref{sec3.3}).In fact, the lowest temperature for optical TDEs that have been discovered is about $\rm \sim 12000\,K$ (e.g., PTF09axc; \citealt{Arcavi2014}), which is comparable to the temperature of ASASSN-18ap around peak, particularly when considering the potential dust reddening as indicated by the over-luminous infrared emission. The late-time temperature from BB-fitting to only a few optical bands yields a significantly lower temperature (i.e. about 5000K) than those from the TDE candidates. However, as discussed in Appendix~\ref{appendiceC}, this result can be significantly contaminated by the host-galaxy light and optical emission lines, and the intrinsic temperature could be high enough for TDE candidates when further considering the potential dust reddening effect. Finally, regarding the possible temperature decline in ASASSN-18ap, it is worth noting that although the radiation temperature of most TDEs remains roughly constant as events gradually fade out, rapid temperature declines have also been observed in a few TDE candidates, such as ASASSN-14ae (\citealt{Holoien2014}) , ASASSN-19bt (\citealt{ASASSN-19bt}),and AT2019qiz(\citealt{AT2019qiz}).

In Figure \ref{spec_comparison}, we compared the different phase spectra of ASASSN-18ap with those of some typical TDEs or TDE candidates. As shown in the figure, very broad emission line ($\rm \sim 10000\,km/s$) was usually detected around the peak, which could be too faint to be detected at early time (e.g., the most early spectra of ASASSN-19bt) and usually becomes narrower at late time (e.g., \citealt{vV2020ssr,Charalampopoulos2022}. However, optical spectroscopic observations of ASASSN-18ap mainly reveal intermediate-width broad Balmer lines with FWHM of $\rm \sim 4000\,km\,s^{-1}$ narrowing to $\rm \sim 2000\,km\,s^{-1}$ at late times. Around the peak, a very broad component with FWHM $\rm 19000\,km\,s^{-1}$ was also detected (see \ref{all_Ha}), which is expected to be detected in tidal disruption events, but at this time the intermediate component still dominates the line profile, opposite to most TDEs (e.g. \citealt{vV2020ssr}). Intermediate-width emission with a FWHM of approximately 1500-3000 km/s has been observed in the TDE ASASSN-14li, even in the early spectra, with an extended wing extending up to 10000 km/s in the blue and red directions (\citealt{Holoien2016}), similar to the case of ASASSN-18ap. However, even the first spectra of ASASSN-14li may have evolved significantly from its peak, as the optical peak was not captured due to a 3-month gap between the initial discovery and the pre-discovery non-detection (\citealt{Holoien2016}). PS16dtm, a canonical TDE candidate in an AGN, also exhibits an intermediate width emission line profile ($\rm FWHM\sim 3000-5000\,km\,s^{-1}$) throughout the phase~\citep{PS16dtm}. 

It should also be noted that the late-time spectra were similar to the two dusty TDE candidate AT2019qiz (\citealt{AT2019qizCL}) and AT2017gge (\citealt{Onori2022,Wang2022ApJL}), with mainly an intermediate width broad $\rm H\alpha$ and possible high-ionization lines (see Figure \ref{spec_comparison}). Therefore, we propose that for these dusty events, a significant portion of the intermediate width component may originate from the rich nuclear ambient gas, which already existed before the occurrence of TDE and is illuminated by the TDE radiation. In this scenario, for ASASSN-18ap, the intermediate component in all the spectra mainly originates from pre-existing gas, and the TDE emission lines were too faint to be detected in the early or late times, with only the spectra around the peak capturing a very broad component. The potential reddening along the line of sight for ASASSN-18ap, as indicated by the lower radiation temperature compared to typical TDEs, would further suppress the central TDE radiation and make the intermediate component dominate the line profile, even in the spectra around the peak. This is in contrast to the other two dusty TDEs AT2017gge and AT2019qiz, for which the high radiation temperature may suggest that not much dust is distributed along the line of sight, and the very broad emission lines from TDE itself can be dominant around the peak. 

Another key feature for ASASSN-18ap is the emergence of high-ionization lines at late times, such as $\rm [Fe\,\textsc{vii}]$, which is generally thought to be a possible signal from gas echoes of intrinsic soft X-ray emission from TDEs and has been used to search for candidate TDEs (e.g., ECLE; \citealt{Wang2012}). The coronal line luminosities of ECLEs are of the order of $\rm 10^{40}\,erg/s$, comparable to those of ASASSN-18ap. These coronal lines have also been observed in some TDEs, such as AT2017gge (\citealt{Onori2022}) and AT2019qiz (\citealt{AT2019qizCL}). We have plotted the coronal line luminosity of the collected TDEs in Figure \ref{coronallines}. Obviously, the coronal lines in TDE candidates can be significantly higher than those in Type IIn supernovae and comparable to that of ASASSN-18ap. Additionally, the TDE scenario can naturally explain the discrepancy between the discovery of strong high-ionization coronal lines and the non-detection of X-rays from quasi-simultaneous Swift observation since the coronal lines echo with years-ago intrinsic emission from TDEs. Delayed X-ray brightening has been found in a handful of optical TDEs (e.g., AT2017gge; \citealt{Onori2022,Wang2022ApJL}, ASASSN-15oi; \citealt{Gezari2017}, OGLE16aaa; \citealt{Kajava2020}, AT2019azh; \citealt{AT2019azh}), which could reach a peak luminosity of about $\rm 10^{43}\,erg\,s^{-1}$ and possess a delay of about years. According to this, it is highly likely that the Swift observations performed for ASASSN-18ap missed the X-ray brightening phase, and the X-ray emission is actually luminous enough to provide the energy needed, as we found in the photoionization simulation. On the other hand, the nondetection of X-rays can be obscured in the reprocessing model of TDEs (e.g., \citealt{LU1997,SQ2009,MS2016,Roth2016,Dai2018}).  Moreover, coronal lines are usually thought to be produced in the inner region of the dust, which would evaporate after the outburst and release iron into the gas. Therefore, the radius of the emission region of the coronal lines can be estimated as $\rm R_{cl} \sim~ c\times t_{delay}\sim10^{18} cm$, where $\rm t_{delay}$ is the delay of the infrared peak with respect to the optical peak, and the result is comparable to the one evaluated by the photoionization simulation in section \ref{sec3.4}.

\subsection{The possibility of AGN activity} 
\label{sec4.3}
\begin{figure}[htb]
\figurenum{14}
\centering
\begin{minipage}{0.5\textwidth}
\centering{\includegraphics[angle=0,width=1.0\textwidth]{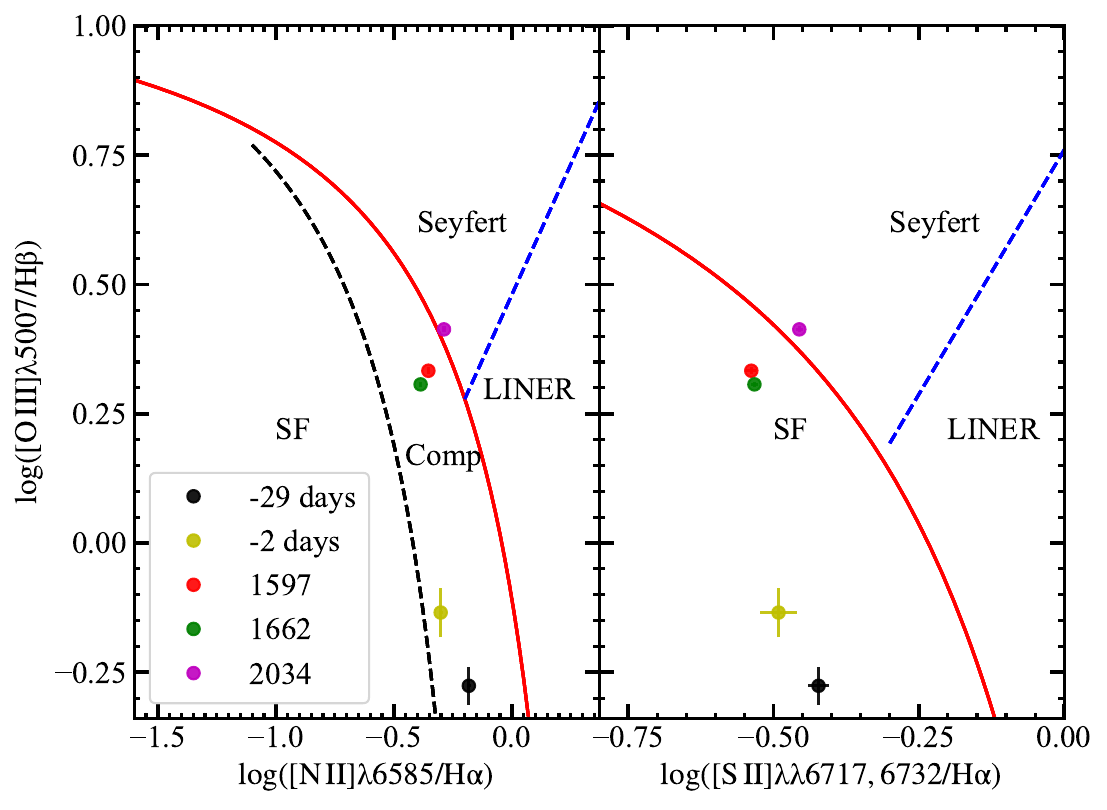}}
\end{minipage}
\caption{ The left panel: The [O\,\textsc{iii}]/$\rm H\beta$ verse [N\,\textsc{ii}]/$\rm H\alpha$ diagnostic diagram. The extreme starburst line by \cite{Kewley2001} and the classification line by \cite{Kauffmann2003} are shown as the red solid and black dashed curves, respectively. The right panel: The [O\,\textsc{iii}]/$\rm H\beta$ verse [S\,\textsc{ii}]/$\rm H\alpha$ diagnostic diagram.\label{BPT}}
\end{figure}

As discussed in Section \ref{sec3.2}, the host SED fitting results of the host and the infrared color disfavor strong AGN activity. Here, we attempted to investigate the AGN activity of the host galaxy with the BPT diagram (\citealt{BPT, VO1987}). As shown in Figure \ref{BPT}, the diagnostic narrow line ratios in both early spectra are located in the composite region near the boundary to the star-forming galaxy in the diagram, indicating a possible weak AGN activity before the occurrence of the transient. Interestingly, the $\rm \frac{[O\,\textsc{iii}]}{H\beta}$ ratio has increased significantly compared to the early-time spectra, moving ASASSN-18ap toward the Seyfert region, and the line ratio in the last spectrum has already entered the Seyfert region. We initially considered the discrepancy between the slit widths of the late-time Palomar/P200 spectra and the two early FLWO/FAST spectra (see Table \ref{specobs}). The narrower slit width of the late-time spectra might result in a higher $\rm \frac{[O\,\textsc{iii}]}{H\beta}$ ratio due to a potential lower contribution from extended starlight. However, all spectra have a similar $\rm [O\,\textsc{ii}$ intensity~\footnote{The spectra around the peak exhibit a significantly higher $\rm [O\,\textsc{ii}$ intensity. However, this is more likely to be a systematic error caused by poor absolute flux calibration, and therefore would not impact the line ratios.} (see Table \ref{spec_line}), indicating a comparable starlight contribution and suggesting that the increase in the $\rm \frac{[O\,\textsc{iii}]}{H\beta}$ ratio is genuine. In fact, similar behavior has been found in the late-time spectra of AT2017gge (\citealt{Onori2022}). Additionally, \citet{Yang2013} reported a significant increase in $\rm [O\,\textsc{iii}]$ in the follow-up spectra of coronal line TDE candidates, shifting the line ratios toward the AGN region on the BPT diagram.

Taking into account the possible existence of weak AGN activity, we further discuss the scenario of AGN activity for ASASSN-18ap. Its substantial brightening amplitude clearly separates ASASSN-18ap from normal AGN variability (\citealt{Vanden2004}). However, a category of extreme AGN variability, known as changing-look (CL) AGN, has been identified and extensively studied in the past decades. These events usually exhibit significant variability amplitudes accompanied by the appearance or disappearance of broad Balmer lines on a timescale of years (e.g. \citealt{KW1971, Macleod2016, Sheng2017}). Although each case may be caused by different reasons ,such as variable obscurer (e.g., \citealt{Goodrich1989}), variability is typically attributed to dramatic change in accretion rate (e.g. \citealt{LaMassa2015, Macleod2016,Sheng2017,Yang2023}). The short rising timescale ( $\rm \sim 40$ days) observed in ASASSN-18ap presents challenges in interpreting it as the accretion state change caused by the disk instability, as the viscosity timescale in a standard thin disk would be orders of magnitude longer (e.g., \citealt{Stern2018}). However, the increasing number of discoveries of extreme AGN variability with short timescales of a year (e.g., \citealt{Rumbaugh2018,Yang2018}) suggests that the standard viscous disk theory may be too simplistic (e.g., \citealt{Lawrence2018}). Thus, interpreting the variability directly using this theory may not be reasonable, and other factors ,such as magnetic fields and outflows, should be considered (e.g., \citealt{Lawrence2018,DB2019,Kaaz2023}). In fact, some models have been developed to account for large-amplitude variability in the optical bands within a relatively short timescale (e.g., \citealt{DB2019}). On the observational side, the rising timescale ( $\rm \gtrsim 100 days$) of extreme AGN variability may be slightly longer than that observed in ASASSN-18ap, such as SDSSJ1115+0544 (\citealt{Yan2019}), and Gaia16aax (\citealt{Cannizzaro2020}). Furthermore, in the changing-look LINER sample of~\cite{Frederick2019}, two of the three sources with rising light curves also exhibit a rising timescale of $\gtrsim 100$~ days, with the fastest rising one, AT2018dyk, still being debated between an AGN or TDE scenario (\citealt{Frederick2019,Huang2023}). However, reaching a conclusion is challenging due to the lack of a well-studied large sample of extreme variability with complete rising light curves. In fact, rapidly rising flares have been observed in narrow-line Seyfert I galaxies, although TDEs have been suggested to be the engine for some of them (e.g., \citealt{Frederick2021}). Finally, the smooth light curve of ASASSN-18ap also disfavors the AGN variability scenario, as small fluctuations in the light curves are typically observed in such a case (e.g. \citealt{Frederick2019}).

TDEs have been proposed as the physical origin of some extreme AGN variability by providing substantial material to the black hole on timescales of months to years (e.g. \citealt{Merloni2015,Trakhtenbrot2019,Zhang2021}). As mentioned in Section \ref{sec4.2}, PS16dtm (\citealt{PS16dtm}), a canonical TDE candidate in an AGN, exhibits emission line features similar to those observed in ASASSN-18ap, such as intermediate-width Balmer lines and late-time coronal lines. Additionally, PS16dtm also displays a slow decay of light curves with a power-law index of $\rm \sim 1$, comparable to that of ASASSN-18ap. Finally, it is worth noting that the location of ASASSN-18ap on the BPT diagram is in the region of a composite galaxy, which also hosts some other TDEs (e.g.\citealt{Hammerstein2021}).

\subsection{Dusty TDEs missed by OUV survey and The MIRONG project} 
\label{sec4.4}
The optical sky survey is currently the fastest and most effective way to search for TDEs, but has a drawback in that it can be affected by dust obscuration, as suggested by the possible misclassification of ASASSN-18ap. Dust grains along the line of sight can alter the transient, making it appear cooler, fainter, or even invisible in optical photometry and spectroscopy. In fact, according to \cite{Jiang2021ApJ}, most optical TDEs discovered before have a low dust-covering factor of about 0.01. Therefore, the optical survey may overlook a significant portion of TDEs in dusty environments, which could be detected through their IR echoes (see also \citealt{Roth2021,Reynolds2022,Panagiotou2023}). To systematically search for TDEs and other nuclear transients in the IR band, \cite{Jiang2021ApJS} constructed a large sample of Mid-IR Outbursts in Nearby Galaxies (MIRONG) by matching SDSS DR14 spectroscopic galaxies with the WISE database. Most sources have been found to have positions close to the galaxy center (median offset <0.1") and have high peak infrared luminosity of about $\rm 10^{43}\,erg\,s^{-1}$, disfavoring the SN scenario. 
Spectroscopic follow-up of MIRONG (\citealt{Wang2022ApJS}) excluded the SN scenario due to the absence of characteristic features of the supernova spectra. Instead, broad $\rm H\alpha$ and coronal lines with similar luminosity and FWHM of ASASSN-18ap and AT2017gge have been discovered in some of the MIRONGs, supporting the TDE origin. Additionally, the integrated energy and peak IR luminosity for the majority sources in the MIRONG sample are comparable with those of the two objects. In fact, AT2017gge was independently discovered by the MIRONG project, while ASASSN-18ap was overlooked only due to the absence of SDSS spectra before the outburst. Therefore, ASASSN-18ap and AT2017gge, as specific cases, support the TDE nature of some MIRONGs, especially those hosted by inactive galaxies.

\subsection{The infrared excess}
\label{sec4.5}
As a potential TDE candidate, the peak IR luminosity of ASASSN-18ap is slightly higher than that of the dusty TDE candidate AT2017gge ($\rm 10^{43.3}\,erg\,s^{-1}$; \citealt{Onori2022,Wang2022ApJL}), which was the most luminous TDE candidate in the IR band discovered in a non-active galaxy before (\citealt{Wang2022ApJL}). Furthermore, the maximum IR luminosity of $\rm 10^{43.51} \,erg\,s^{-1}$ is higher than in the OUV. Interestingly, the integrated IR energy $\rm E_{IR}\sim3.1\times10^{51}\,erg$ is about an order of magnitude higher than that of OUV $\rm E_{opt}\sim3.9\times10^{50}\,erg$, indicating an IR excess. Although there may be a contribution of IR energy from X-ray reprocessing, as evidenced by the high-ionization coronal lines, the observed energy in X-rays for optical TDEs would usually not exceed that of OUV (\citealt{AT2019azh,Kajava2020,Onori2022,Wang2022ApJL}) and thus could not account for the excess. Instead, the infrared excess can be understood in three ways: (i) heavy dust extinction along the line of sight, (ii) EUV (extreme ultraviolet ) bump, or (iii) anisotropic radiation. In the first case, as detailed in Appendix \ref{appendiceB}, the OUV energy can reach the IR level when an extra dust extinction from the host galaxy is considered, with $\rm A_v\sim0.97$ and a dust extinction curve of the Small Magellanic Cloud (\citealt{Gordon2003}). Furthermore, the radiation temperature around peak would be $\rm \sim 20000\,K$, closer to the typical value for TDEs. In this scenario, the intrinsic X-ray emission may also suffer significant absorption by the dense gas along the line of sight, making it too faint to be detected by Swift/XRT. In the second case, the bolometric luminosity is dominated by strong EUV radiation. In particular, an EUV scenario has also been proposed to address the missing energy problem of TDEs, where the observed energy is one to two magnitudes lower than theoretically predicted (e.g., \citealt{Lu2018}).  Finally, anisotropic radiation may simultaneously explain the IR excess, the slightly lower temperature than the typical value of TDEs, and also the non-detection of X-rays if the source is observed edge-on in the unified model of \cite{Dai2018}. 

\subsection{The early bump in the optical light curve} 
\label{sec4.6}
An obvious bump is discovered at the onset of ASASSN-18ap brightening in the ATLAS-o band (see Figures \ref{SN2018gnlc} and \ref{wholeLC}), which has not been observed in any TDE before, except for a faint candidate TDE AT2020wey (\citealt{AT2020wey}) and the TDE candidate ASASSN-19bt hosted by a Seyfert galaxy (\citealt{ASASSN-19bt}). However, the precursor observed in AT2020wey is too weak to be reliably recognized and to provide further insight, due to the low cadence and poor data quality. Furthermore, while for ASASSN-19bt the bump feature was detected in the bolometric luminosity light curve derived from the ASASSN-g band light curve using temperature from Swift/UVOT SEDs, the bump was not directly detected in the ASASSN-g band. For ASASSN-18ap, the bump feature is much more reliable, with a luminosity of the order of $\rm 10^{42}\,erg\,s^{-1}$ and a timescale of $\rm \sim 10~days$. Due to the lack of understanding of the emission mechanism driving the light-curve rise of TDEs, it is challenging to directly determine the physical process behind the early bump, and detailed analysis, theoretical work, or simulations are necessary to explore the physical origin. In fact, such a feature could be caused by radiation from electron recombination and cooling in unbound debris (e.g., \citealt{Kasen2010}), stream-stream collision (e.g., \citealt{Kim1999,Piran2015}), wind-stream collision (e.g., \citealt{Calder2024}), and vertical shock compression during the first passage (e.g., \citealt{Yalinewich2019}). In addition, we note that the stream motion is highly hypersonic. A strong shock is produced when two streams collide and propagate to the back sides.  Thus, a shock breakout, similar to those in SNe, may be expected when the shock reaches the back surface of the streams before the expansion of the collided debris. In analogy to the SN shock break, the total energy radiated can be estimated $\rm E_{bo}\simeq 2.2\times 10^{49}\ R_{14}^2v_{bo,9}\kappa_{0.34}^{-1}\ erg$ where the debris size is $\rm R_{14}10^{14}$ cm, the velocity of the shock $\rm v_{bo}=10^9v_{bo,9}\ cm\,s^{-1}$ and the opacity $\rm \kappa=0.34 \kappa_{0.34}\ cm^2\,g^{-1}$ (e.g.\citealt{Katz2012}). 

\begin{figure*}[htb]
\figurenum{15}
\centering
\begin{minipage}{1.0\textwidth}
\centering{\includegraphics[angle=0,width=1.0\textwidth]{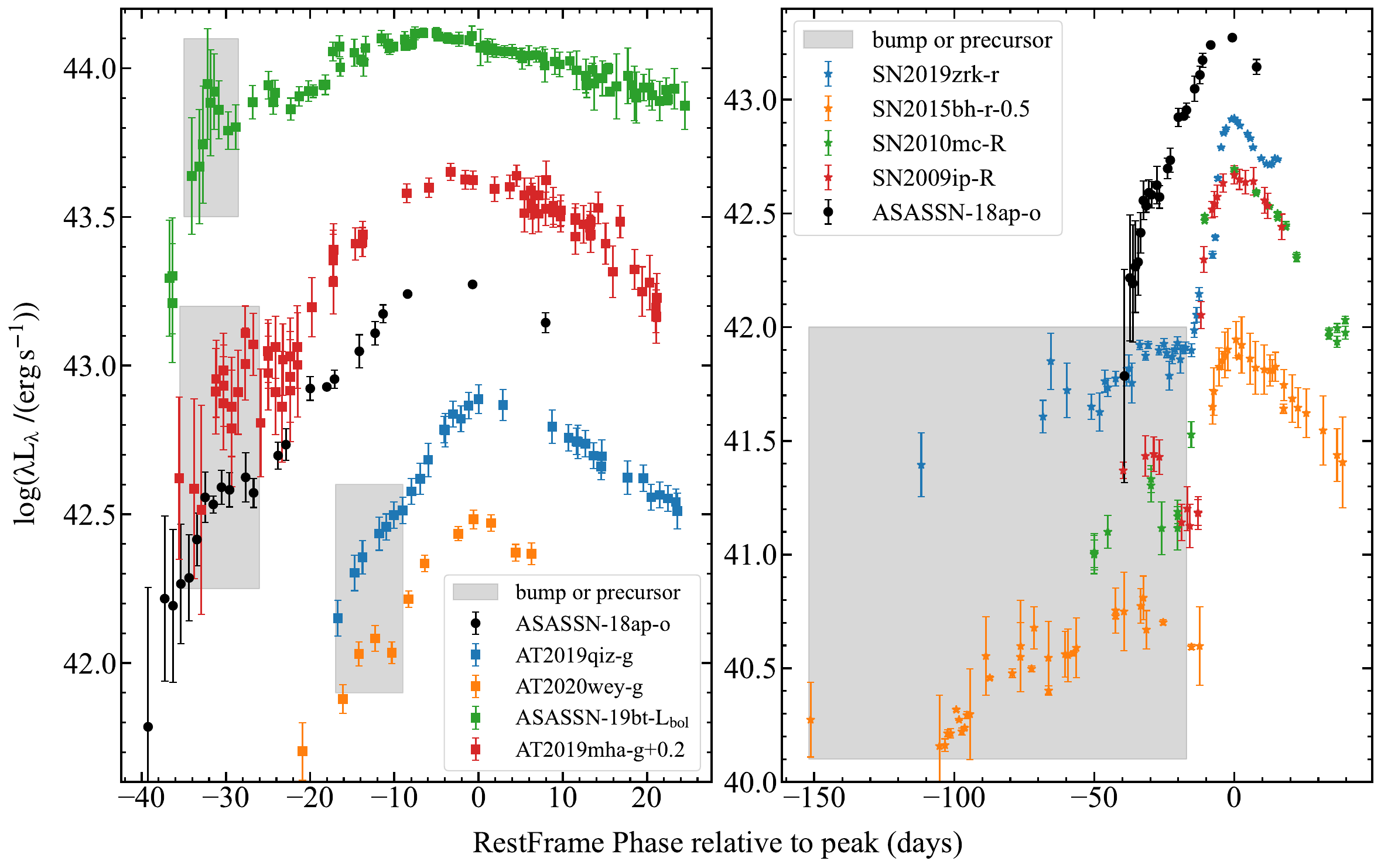}}
\end{minipage}
\caption{ The left panel presents the rising light curve of ASASSN-18ap, along with TDE candidates that exhibit an early bump, including AT2020wey (\citealt{AT2020wey}), ASASSN-19bt (\citealt{ASASSN-19bt}), AT2019qiz (e.g., \citealt{AT2019qiz}), and AT2019mha. The bump feature of the latter two sources was first reported in this work. To improve clarity, we have shifted the rising light curve of AT2019mha by +0.2 dex. The right panel displays some precursors found in Type IIn supernovae: SN2009ip (\citealt{SN2009ip}), SN2010mc (\citealt{SN2010mc}), SN2015bh (\citealt{SN2015bh}), and SN2019zrk (\citealt{SN2019zrk}). Also, the light curve of SN2015bh was shifted by -0.5dex for clarity. \label{bumpComparison}}
\end{figure*}

Encouraged by the discovery of early bumps in ASASSN-18ap, ASASSN-19bt, and AT2020wey, we immediately examined the ZTF light curves of the 30 TDEs from the ZTF-I survey (\citealt{Hammerstein2023}) and identified two more sources, AT2019mha and AT2019qiz, both of which exhibit a similar and reliable bump feature at the onset of the rising phase. In Figure \ref{bumpComparison}, we present all five sources with early bumps. Although the luminosity around the bump for these sources spans a wide range, they demonstrate a similar relative intensity to the peak luminosity, i.e. a fraction of the peak luminosity. Additionally, the timescale of these sources was around 10 days. Notably, AT2019mha displays a rising light curve with the same shape as ASASSN-18ap, although they decay following a different slope. These findings also suggest that ASASSN-18ap is a candidate for TDE, as a similar early bump structure has not been observed in Type IIn supernovae (more details are discussed below). Interestingly, all five targets show a decrease in temperature (see Figure \ref{LC_comparison}) around the peak, except AT2019mha, which lacks multiband photometry. Since a significant fraction of ZTF-I TDEs do not have a complete or high-quality light curve for the rising phase, the number of TDEs with an early bump should be higher. In fact, the ZTF light curves of AT2019qiz do not provide a reliable detection of the early bump, and the additional LCO data (from \citealt{AT2019qiz}) help to confirm this. The upcoming dedicated optical time-domain surveys, such as the Legacy Survey of Space and Time (LSST; \citealt{Ivezic2019}) and the deep high-cadence survey of Wide Field Survey Telescope (WFST; \citealt{Lin2022,Wang2023}), will enable us to characterize the early rising light curves much more accurately and to determine the occurrence frequency of the early bump. Investigating the physical process leading to the bump feature may provide insight into the emission mechanism driving the rise of the TDE light curves.  

In the Type IIn scenario, pre-SN outbursts with intermediate luminosity associated with episodic ejection of matter have been observed weeks to years prior to the terminal explosion (e.g., \citealt{Ofek2014,Strotjohann2021}), such as SN2009ip (e.g., \citealt{SN2009ip}), SN2010mc (\cite{SN2010mc}), SN2015bh (\citealt{SN2015bh}), and SN2019zrk (\citealt{SN2019zrk}). These precursor eruptions are commonly known as SN imposters (e.g. \citealt{Filippenko1995,VanDyk2000}), and the triggering mechanism is still not well understood (e.g. \citealt{Langer2012,Smith2014}). In Figure \ref{bumpComparison}, we compare the early bump structure of ASASSN-18ap with some well-studied events featuring such precursors. It is evident that these Type IIn precursors possess a longer timescale of months to years and lower luminosity ($\rm \lesssim 10^{42}\,erg\,s^{-1}$). In contrast to these independent precursors leading to the main outburst, ASASSN-18ap's bump appears more like an excess in flux at the onset of the brightening of the main outburst.
On the other hand, the earliest emission of supernovae was thought to be produced by the shock breakout, which occurs when the shock reaches the edge of the supernova, and the optical depth drops below $\rm c/v_{shock}$ (e.g. \citealt{Katz2012, WK2017}). This breakout is luminous in X-ray/UV, with a timescale of seconds to a fraction of an hour, two orders of magnitude lower than ASASSN-18ap (e.g. \citealt{WK2017}). Although the post-shock cooling emission could produce a UV/optical flare lasting several days and sometimes combined with radioactive emission form a double-peak light curve, the double-peak light curve is inconsistent with the bump feature and is mainly detected in Type IIb supernovae (e.g., \citealt{Arcavi2011,Arcavi2017,WK2017,Bersten2018}). Additionally, we note that a similar excess in flux at the early light curve has been observed in Type Ia supernovae (e.g., \citealt{SN2018oh,ASASSN-18bt}), but the Type IIn scenario should be the only possible supernova scenario for ASASSN-18ap.

\section{Conclusion} 
\label{sec5}
In this work, we have revisited the classification of the transient ASASSN-18ap by analyzing its long-term multiband light curves and spectroscopic evolution, incorporating newly acquired late-time spectra. Both the TDE and Type IIn supernova scenarios roughly conform to the optical photometric and spectroscopic observations, although the low temperature and weakness of the very broad component may indicate a potential reddening effect along the line of sight in the TDE scenario. Based on the high infrared luminosity and the strong narrow lines that emerged at late times, we are inclined to classify ASASSN-18ap as a plausible TDE candidate, although an extraordinary Type IIn cannot be excluded. Interestingly, a bump has been found at the onset of brightening, which is similar to that found in TDEs. More efforts are needed to explore the frequency of this characteristic among the TDE population and its physical origin. Below, we summarize the properties of ASASSN-18ap:

\begin{enumerate}
\item[$\bullet$] ASASSN-18ap reaches a maximum luminosity of $\rm 10^{43.41}\,erg\,s^{-1}$ in 40 days and then decays according to a power-law with index $\rm \sim -0.75$. The temperature of ASASSN-18ap around its peak was approximately $\rm \sim 10000\,K$ and remained nearly constant for at least 10 days. The late-time temperature may have decreased, but a reliable value could not be obtained due to the absence of UV photometry. The total integrated energy of the OUV is about $\rm 3.9\times10^{50}\,erg$. These properties conform well to those of canonical TDEs except for a lower temperature, which might be caused by dust extinction on the light of sight, as indicated by its over-luminous infrared emission.  

\item[$\bullet$] At $\rm \sim 500$ days after the optical peak, ASASSN-18ap reaches its peak in infrared light curves with a luminosity of $\rm 10^{43.51}\,erg$. To date the integrated infrared energy is $\rm \sim 3.1\times 10^{51}\,erg$, one order of magnitude higher than the optical one (called infrared excess in this paper). The infrared luminosity and integrated energy are higher than that of the most luminous supernova in IR ever found, and thus largely exclude the SN scenario for ASASSN-18ap. The infrared excess indicates either heavy dust extinction along the sight or the presence of an additional EUV component in its SED or highly anisotropic emission.

\item[$\bullet$] The key spectroscopic features of ASASSN-18ap include an intermediate-width $\rm H\alpha$ with FWHM $\rm \sim 2000-4000\,km\,s^{-1}$, a very broad $\rm H\alpha$ with FWHM $\rm \sim19000\,km/s$ , and the presence of high-ionization lines with luminosity $\rm \sim 10^{40}\,erg\,s^{-1}$ at late-time. The intermediate component was dominant throughout the entire duration of the transient, while the very broad one was only detected around the peak. This is incompatible with typical TDEs, but could be partly interpreted by the potential presence of rich dust and gas surrounding the nuclei. While these spectroscopic features align with typical Type IIn SNe, the coronal line luminosities exceeded that observed in Type IIn events by at least one order of magnitude. Photoionization simulations using CLOUDY yielded a significantly larger emitting radius than expected from supernova ejecta expansion, as well as a markedly higher X-ray luminosity compared to typical Type IIn events.

\item[$\bullet$] ASASSN-18ap exhibits a clear bump at the onset of brightening in the optical light curve, which has not been discovered in previous TDEs except for AT2020wey and ASASSN-19bt. We checked ZTF light curves of the 30 TDEs from ZTF-I survey artificially,  and found a similar bump in AT2019mha and AT2019qiz. The bump feature in these events, show a similar timescale of 10 days, as well as a similar relative intensity to the peak luminosity. More research is needed to understand the physical process behind this characteristic. 
\end{enumerate}

\acknowledgements
This work is supported by the National Key R\&D Program of China (No. 2023YFA1608100), the SKA Fast Radio Burst and High Energy Transients Project (2022SKA0130102), the National Natural Science Foundation of China (grants 12393814, 12073025, 12103048, 12192221, 12233008), the Strategic Priority Research Program of the Chinese Academy of Sciences (XDB0550200, XDB0550202) and the China Manned Space Project (No.CMS-CSST-2021-A12). The authors acknowledge the support of the Cyrus Chun Ying Tang Foundations. This research uses data obtained through the Telescope Access Program (TAP). Observations with the Hale Telescope at Palomar Observatory were obtained as part of an agreement between the National Astronomical Observatories, the Chinese Academy of Sciences, and the California Institute of Technology. 

\clearpage
\restartappendixnumbering
\appendix
\onecolumngrid
\section{Spectroscopic fitting and results} 
\label{appendiceA}
\begin{figure*}[htb]
\centering
\begin{minipage}{1.0\textwidth}
\centering{\includegraphics[angle=0,width=1.0\textwidth]{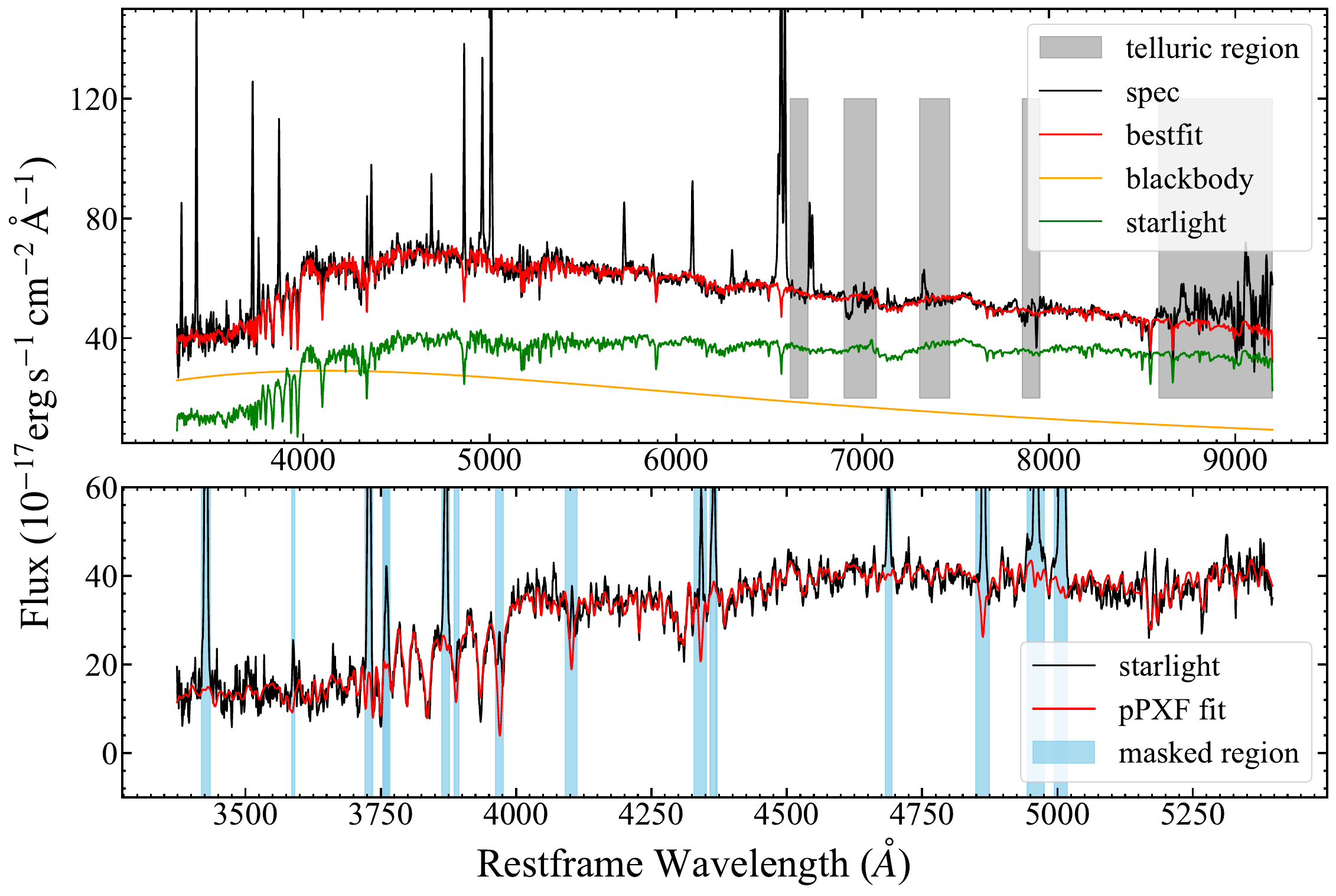}}
\end{minipage}
\caption{ As an example, we present the continuum, velocity dispersion fitting of the spectrum taken on UT2022-09-02 with P200/DBSP. (a) The top panel shows the results of the continuum fitting. The black curve represents the original spectrum data after Galactic extinction correction and redshift correction, while the red curve shows the best-fit results obtained by combining starlight components (green curve) and a blackbody component (orange curve). The gray-shaded region indicates the telluric regions, which were masked along with other apparent emission line regions during the fitting process. (b) The second panel shows the pPXF fitting to the starlight components that we acquired in the continuum fitting process, and the light-blue-shaded region marked the emission line region which was masked in the fitting process.\label{spec_fit}}
\end{figure*}

\begin{figure*}[htb]
\centering
\begin{minipage}{1.0\textwidth}
\centering{\includegraphics[angle=0,width=1.0\textwidth]{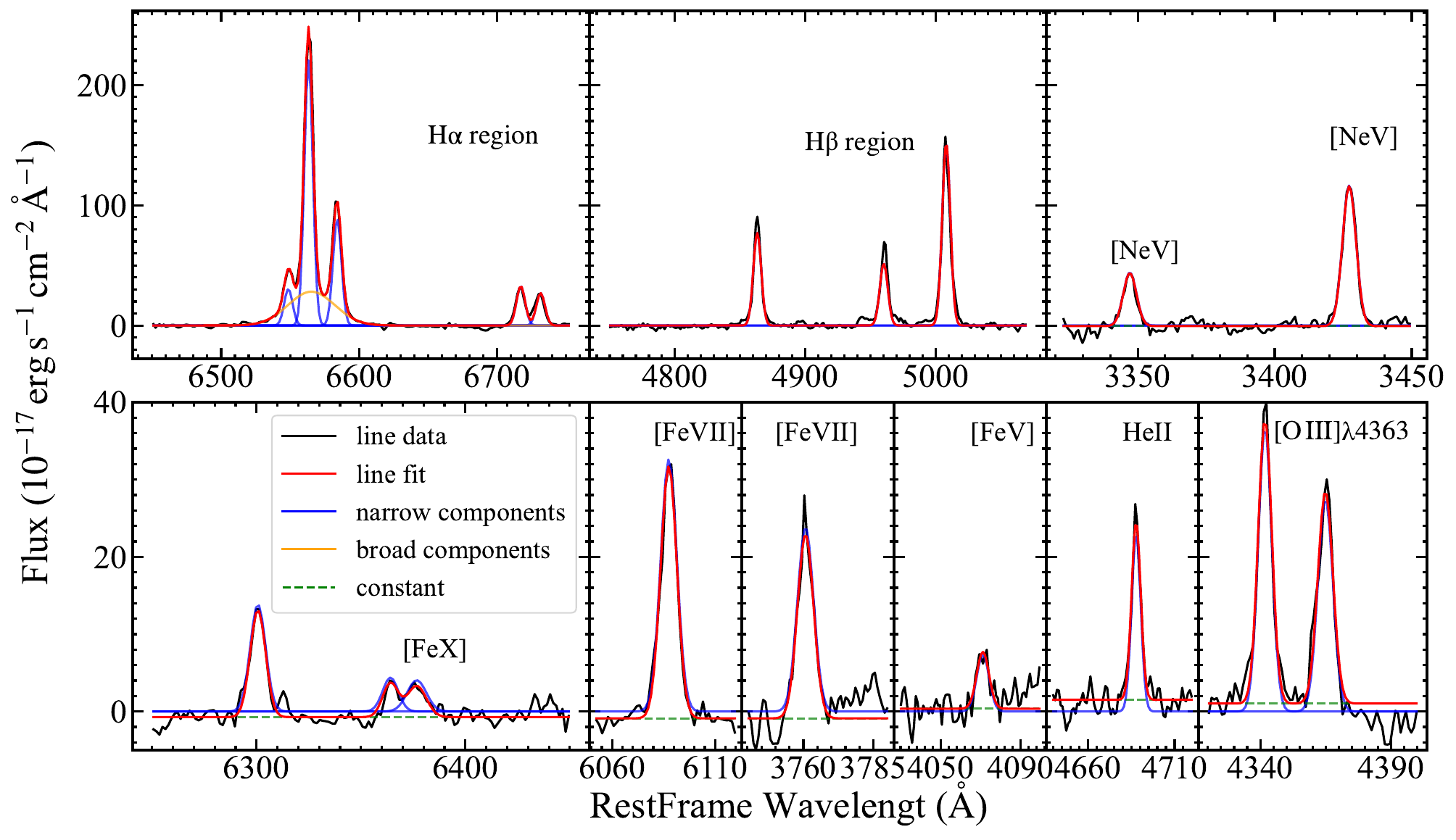}}
\end{minipage}
\caption{ The emission line fitting to some characteristic emission lines in the spectrum taken on UT2022-09-02. The black, red, blue, and orange curves represent the line data, best fit, narrow Gaussian components, and broad Gaussian components, respectively. A constant was added in some emission line regions to correct the local continuum residuals\label{line_fit}}
\end{figure*}

As an example of the spectral analysis mentioned in Section \ref{sec3.1}, we present the continuum and emission line fitting of the spectrum taken on 2022-09-02 in Figure \ref{spec_fit} and \ref{line_fit} respectively, in this appendix. Additionally, we provide the emission line properties derived from the five spectra of ASASSN-18ap in Table \ref{spec_line}.

\tabletypesize{\footnotesize}
\begin{deluxetable*}{ccccccccccc}[htb]
\setlength{\tabcolsep}{0.06in}
\tablecaption{ \it characteristic lines \label{spec_line}}
\tablewidth{0pt}
\tablehead{    & \multicolumn{2}{c}{FAST-2018-01-15}   & \multicolumn{2}{c}{FAST-2018-02-11\tablenotemark{\footnotesize$\rm 1$} } 
               & \multicolumn{2}{c}{P200-2022-06-29}   & \multicolumn{2}{c}{P200-2022-09-02}  
               & \multicolumn{2}{c}{P200-2023-09-09}  \\
Emission-line  &   flux&FWHM                           &  flux&FWHM       
               &   flux&FWHM                           &  flux&FWHM 
               &   flux&FWHM                          \\
               & $\rm 10^{-17} erg\,s^{-1}\,cm^{-2}$&km/s  & $\rm 10^{-17} erg\,s^{-1}\,cm^{-2}$&km/s 
               & $\rm 10^{-17} erg\,s^{-1}\,cm^{-2}$&km/s  & $\rm 10^{-17} erg\,s^{-1}\,cm^{-2}$&km/s 
               & $\rm 10^{-17} erg\,s^{-1}\,cm^{-2}$&km/s }
\startdata
$\rm H\,\alpha$&1358.23$\pm$23.47&  309$\pm$4&4162.83$\pm$85.84&  296$\pm$5&1553.77$\pm$13.36&  406$\pm$2&1677.85$\pm$7.28&  330$\pm$1&1466.31$\pm$8.94&  355$\pm$1\\
$\rm H\,\alpha_{I}$\tablenotemark{\footnotesize$\rm 2$}&2822.66$\pm$75.12&  3958$\pm$130&12982.80$\pm$356.53&  3409$\pm$82&904.28$\pm$25.20&  2020$\pm$36&1143.08$\pm$13.13&  1794$\pm$16&1011.36$\pm$16.32&  1665$\pm$20\\
$\rm H\,\alpha_{B}$\tablenotemark{\footnotesize$\rm 3$}&-----            &  -----       & 13944 $\pm$1000   &  19074$\pm$1327 &  -----      &  -----     & -----   & ----- &----- &-----\\
$\rm [O\,\textsc{iii}]\lambda5007$&209.28$\pm$13.28&  474$\pm$21 &562.96$\pm$40.99&  346$\pm$22&1161.26$\pm$10.37&  439$\pm$3&1183.98$\pm$7.05&  439$\pm$2&1562.55$\pm$8.48&  453$\pm$2\\
$\rm [O\,\textsc{iii}]\lambda4959$&70.23$\pm$4.46&  474$\pm$21&188.91$\pm$13.76&  346$\pm$22&389.69$\pm$3.48&  439$\pm$3&397.31$\pm$2.37&  439$\pm$2&524.35$\pm$2.85&  453$\pm$2\\
$\rm [O\,\textsc{iii}]\lambda4363$&-----&-----&-----&-----&177.10$\pm$8.61&  425$\pm$21&198.52$\pm$6.34&  466$\pm$15&238.35$\pm$6.97&  470$\pm$14\\
$\rm H\,\beta $&393.15$\pm$22.37&  474$\pm$21&767.18$\pm$63.15&  346$\pm$22&539.25$\pm$7.27&  439$\pm$3&584.28$\pm$6.12&  439$\pm$2&603.53$\pm$6.99&  453$\pm$2\\
$\rm H\,\beta_{I} $&552.82$\pm$52.56&  3647$\pm$401&3698.84$\pm$162.86&  2983$\pm$156&-----&-----&-----&-----&-----&-----\\
$\rm [O\,\textsc{ii}]\lambda3728$&690.30$\pm$50.68&  608$\pm$46&1327.49$\pm$143.32&  408$\pm$51&531.05$\pm$10.30&  583$\pm$11&583.88$\pm$6.89&  546$\pm$6&614.77$\pm$8.45&  558$\pm$7\\
$\rm [Fe\,\textsc{x}]\lambda6374.5$&-----&-----&-----&-----&29.33$\pm$4.52&  333$\pm$56&46.26$\pm$3.76&  505$\pm$45&19.47$\pm$4.33&  432$\pm$109\\
$\rm [Fe\,\textsc{vii}]\lambda5159$&-----&-----&-----&-----&33.31$\pm$8.34&  441$\pm$48&46.77$\pm$6.00&  434$\pm$33&47.13$\pm$5.39&  310$\pm$21\\
$\rm [Fe\,\textsc{xiv}]\lambda5304$&-----&-----&-----&-----&13.53$\pm$7.59&  73$\pm$36&-----&-----&38.58$\pm$6.13&  355$\pm$62\\
$\rm He\,\textsc{i}\lambda5876$&-----&-----&-----&-----&79.69$\pm$5.17&  479$\pm$34&65.78$\pm$4.90&  530$\pm$42&72.08$\pm$4.30&  446$\pm$28\\
$\rm He\,\textsc{ii}\lambda4686$&-----&-----&-----&-----&149.50$\pm$7.15&  378$\pm$19&151.65$\pm$5.86&  396$\pm$16&101.92$\pm$6.71&  354$\pm$24\\
$\rm [Ar\,\textsc{iii}]\lambda7136$&-----&-----&-----&-----&35.79$\pm$4.02&  303$\pm$34&54.15$\pm$3.51&  338$\pm$22&65.86$\pm$3.28&  342$\pm$17\\
$\rm [Fe\,\textsc{vii}]\lambda6087$&-----&-----&-----&-----&304.54$\pm$6.16&  457$\pm$9&322.29$\pm$4.90&  457$\pm$6&256.86$\pm$4.77&  479$\pm$9\\
$\rm [Fe\,\textsc{vii}]\lambda5721$&-----&-----&-----&-----&197.16$\pm$6.15&  444$\pm$14&225.60$\pm$5.82&  525$\pm$13&167.18$\pm$4.99&  453$\pm$13\\
$\rm [Fe\,\textsc{vii}]\lambda3759$&-----&-----&-----&-----&138.23$\pm$8.49&  473$\pm$30&172.09$\pm$6.26&  541$\pm$20&142.78$\pm$7.90&  618$\pm$35\\
$\rm [Fe\,\textsc{vii}]\lambda3586$&-----&-----&-----&-----&147.23$\pm$16.17&  877$\pm$110&69.24$\pm$5.32&  368$\pm$29&77.22$\pm$7.54&  469$\pm$50\\
$\rm [Fe\,\textsc{vi}]\lambda5176$&-----&-----&-----&-----&64.58$\pm$9.04&  441$\pm$48&67.71$\pm$6.39&  434$\pm$33&57.04$\pm$5.50&  310$\pm$21\\
$\rm [Fe\,\textsc{vi}]\lambda5146$&-----&-----&-----&-----&17.60$\pm$7.84&  441$\pm$48&34.47$\pm$5.33&  434$\pm$33&31.04$\pm$4.71&  310$\pm$21\\
$\rm [Fe\,\textsc{v}]\lambda4071$&-----&-----&-----&-----&59.98$\pm$8.20&  441$\pm$60&56.62$\pm$6.34&  534$\pm$61&57.34$\pm$6.58&  443$\pm$53\\
$\rm [Ne\,\textsc{v}]\lambda3346$&-----&-----&-----&-----&238.50$\pm$5.44&  457$\pm$10&266.97$\pm$3.54&  509$\pm$7&257.54$\pm$4.31&  563$\pm$9\\
$\rm [Ne\,\textsc{v}]\lambda3426$&-----&-----&-----&-----&644.77$\pm$14.71&  457$\pm$10&721.72$\pm$9.57&  509$\pm$7&696.24$\pm$11.65&  563$\pm$9\\
$\rm [Ne\,\textsc{iii}]\lambda3869$&-----&-----&-----&-----&358.21$\pm$7.77&  414$\pm$9&368.27$\pm$5.19&  463$\pm$6&464.40$\pm$6.54&  449$\pm$6\\
$\rm [Ne\,\textsc{iii}]\lambda3967$&-----&-----&-----&-----&108.40$\pm$2.35&  414$\pm$9&111.45$\pm$1.57&  463$\pm$6&140.54$\pm$1.98&  449$\pm$6\\
$\rm [Fe\,\textsc{xiv}]\lambda5304$&-----&-----&-----&-----&13.53$\pm$7.59&  73$\pm$36&-----&-----&38.58$\pm$6.13&  355$\pm$62\\
\enddata
\tablecomments{ We list some characteristic emission line fitting results for the five spectra of ASASSN-18ap. The error of emission line flux in this table is the statistical error given by the emission line fitting procedure outlined in Section \ref{sec3.1}. } 
\tablenotetext{1}{The spectrum taken around peak have systematically stronger narrow lines than the others. This could be caused by a poor flux calibration. }
\tablenotetext{2}{The subscript 'I' represents the intermediate width component}
\tablenotetext{3}{The subscript 'B' represents the very broad component}
\end{deluxetable*}

\clearpage
\onecolumngrid
\section{Dust extinction from the Host galaxy} 
\label{appendiceB}
\begin{figure*}[htb]
\centering
\begin{minipage}{1.0\textwidth}
\centering{\includegraphics[angle=0,width=1.0\textwidth]{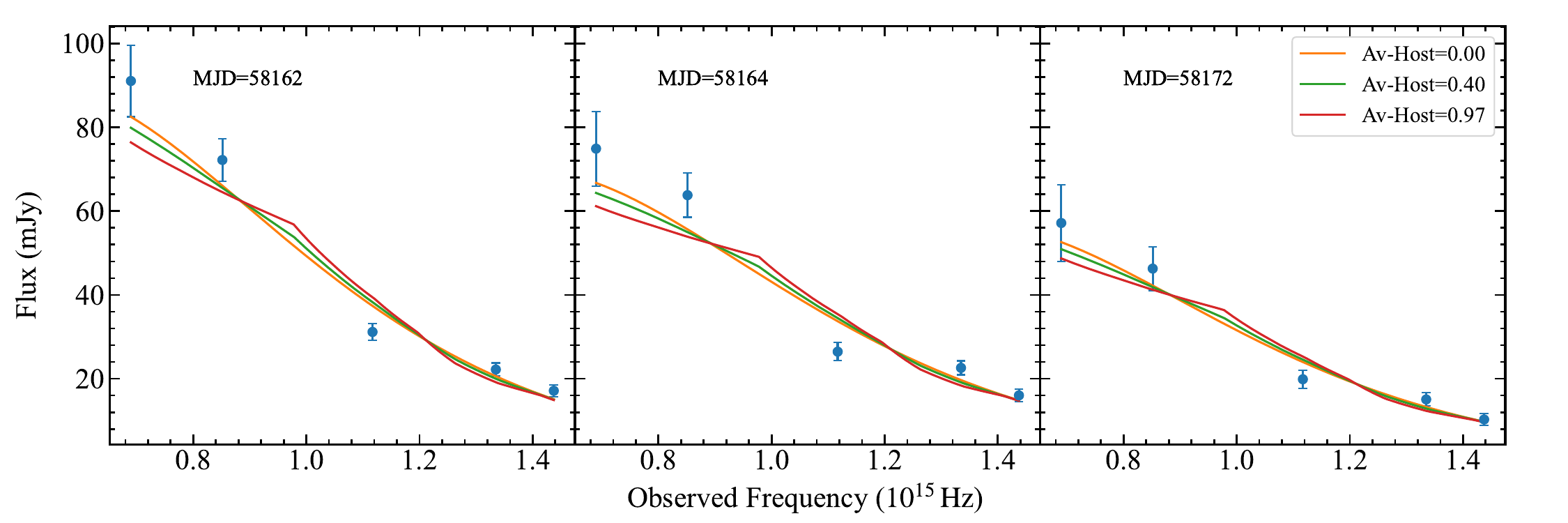}}
\end{minipage}
\caption{We show the black-body fitting to the multi-band Swift/UVOT photometry (blue points) at three epochs, with various levels of host galaxy extinction, including $\rm A_{v}=0.0$ (orange curve), $\rm A_{v}=0.4$ (green curve) and $\rm A_{v}=0.97$ (red curve).\label{hostextinction}}
\end{figure*}

In comparison to other TDEs, ASASSN-18ap has a slightly lower temperature but a higher infrared luminosity, possibly attributed to a significant amount of dust extinction along the line of sight from the host galaxy. To investigate this, we used an extinction curve from the Small Magellanic Cloud (\cite{Gordon2003}) and compared the black-body fitting results with the Swift/UVOT photometry corrected for Galactic extinction with different levels of host-galaxy extinction, including $\rm A_{v}=0.0$, $\rm A_{v}=0.4$, and $\rm A_{v}=0.97$. The multi-band photometry SEDs of ASASSN-18ap were roughly fitted by all three models, as illustrated in Figure \ref{hostextinction}. Although the outcome with no host-galaxy extinction (i.e. $\rm A_{v}=0.0$) yielded the lowest $\chi^{2}$ value,  it did not exhibit statistical superiority over the other results at an 80\% confidence level, based on the F distribution with both degrees of freedom being 3. Consequently, a significant host extinction cannot be ruled out, allowing the potential application of high host extinction to elevate the intrinsic radiation temperature and integrated energy. If $\rm A_{v}=0.97$, the black-body fitting would produce a temperature of $\rm \sim 20000,K$ and an integrated OUV energy of $\rm \sim 2.5\times 10^{51}\,erg$~\footnote{ We utilized the bolometric light curve generated by scaling the monochromatic light curve (refer to section \ref{sec3.3}), as the late-time optical photometry can not distinctly demonstrate the extinction effect. } for ASASSN-18ap. Finally, based on the extinction law in \cite{WangChen2019}, dust extinction has a negligible effect on the infrared bands of W1 and W2.


\section{The reliability of blackbody fitting results without UV photometry} 
\label{appendiceC}
\begin{figure*}[htb]
\centering
\begin{minipage}{1.0\textwidth}
\centering{\includegraphics[angle=0,width=1.0\textwidth]{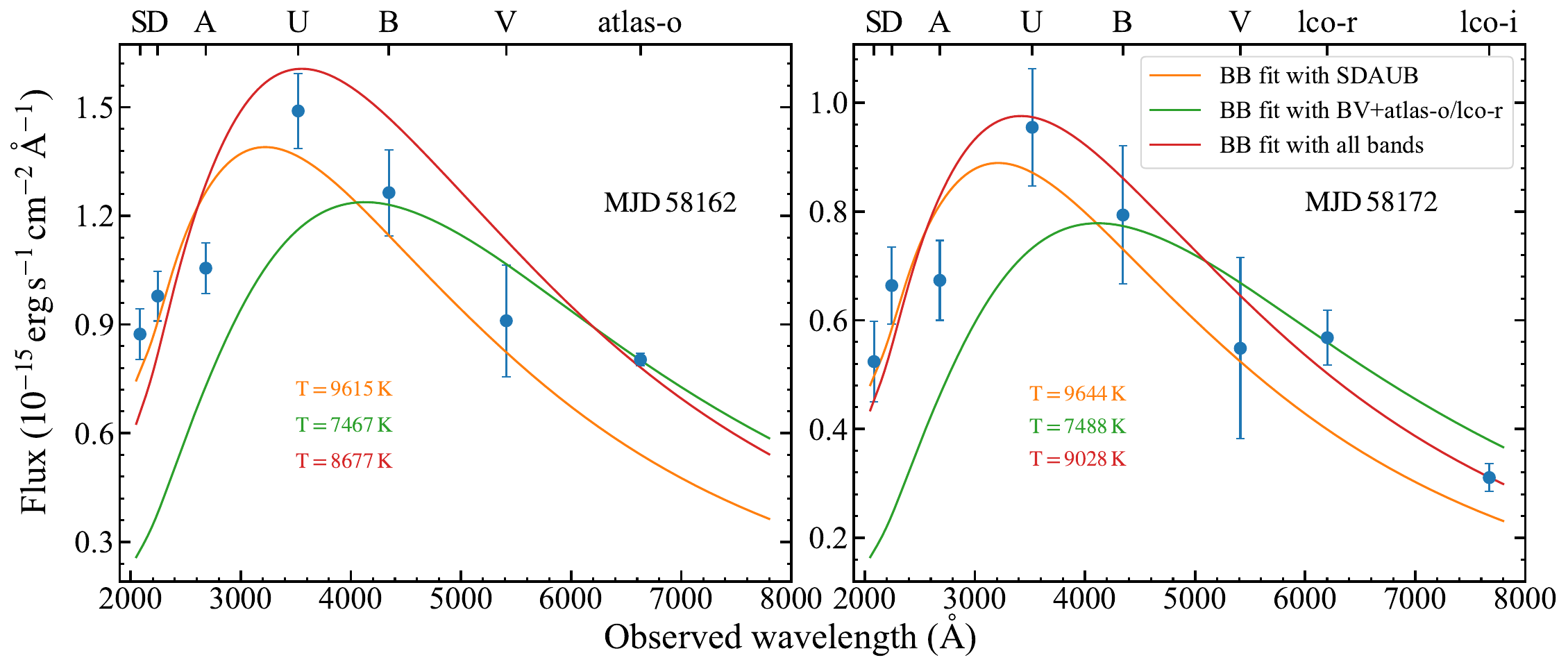}}
\end{minipage}
\caption{ Here we present the blackbody fitting results to $\rm MJD\,58162$ (left panel) and $\rm MJD\,58172$ (right panel), for three cases with different bands included. The orange, green and red curves represent the results of case i, ii, iii mentioned in the text respectively. The temperatures of the three cases were labeled with the same color code as the corresponding fitting curve.  \label{bbfit-bands}}
\end{figure*}

In section~\ref{sec3.3}, we employed blackbody (BB) fitting at late-time with the five bands: ATLAS-o, ATLAS-c, ZTF-g, ZTF-r, Gaia-G. However, all of these bands are easily contaminated by the host galaxy light, and therefore potentially underestimate the blackbody temperature. To illustrate this, we conducted blackbody fitting with different bands included, using the multi-band photometry obtained around the peak at MJD 58162 and MJD 58172. Both datasets contained Swift/UVOT photometry and ATLAS-o (or LCO-r/LCO-i) band data within a day. Specifically, we considered three cases: (i) BB fit with Swift/UVOT photometry except for the V-band (ii) BB fit with the Swift-B, Swift-V, and ATLAS-o (or LCO-r), since these three bands are similar to the ATLAS-o, ZTF-g, ZTF-r bands with the highest cadence for ASASSN-18ap at late-time (iii) BB fit with all the bands. As shown in the figure~\ref{bbfit-bands}, the BB fitting results with only three optical bands (similar to the BB fitting for ASASSN-18ap at late-time) yielded a significantly lower temperature as expected, about $\rm 2000\,K$ lower than the one with five Swift/UVOT bands and $\rm 1000-1500\,K$ lower than the one with all bands. Interestingly, the LCO-r band or ATLAS-o bands deviated the most from the entire blackbody spectral energy distribution, and this could be caused by the coverage of the most prominent emission lines $\rm H\alpha$ in these two bands. On the other band, despite the lower temperature derived from BB fitting with only three optical bands, approximately $\rm 7500\,K$, it was still higher than the one obtained for ASASSN-18ap at late-times (about $\rm5000\,K$, see \ref{sec3.3}), which makes the decline in temperature after peak more convincing. However, it can not be confirmed since the evolution of host-galaxy contamination and emission lines influence was unknown.

\clearpage
\twocolumngrid
\section{Swift/XRT data reduction and host SED }
\label{appendiceD}
In this appendix, we show the results of the Swift / XRT data analysis in Table \ref{swift/XRT}, and the host SED collected from several archives in Table \ref{host-photos}. 

\begin{deluxetable}{cccccccc}[htb]
\tabletypesize{\footnotesize}
\setlength{\tabcolsep}{0.06in}
\tablecaption{ \it Swift/XRT upper limits \label{swift/XRT}}
\tablewidth{0pt}
\tablehead{MJD  & exposure &$\rm cts_{src}$ & $\rm cts_{bkg}$  &  $\rm cts_{upper}$ & $\rm ctr_{upper}$    & $\rm fTDE_{unabs}$ &  $\rm fSNe_{unabs}$   \\
            &  s       & &  &  &  cts/s  & $\rm 10^{-12} erg\,s^{-1}\,cm^{-2} $  \\
        \colhead{(1)} &\colhead{(2)}&\colhead{(3)}&\colhead{(4)}&\colhead{(5)}&\colhead{(6)}&\colhead{(7)}&\colhead{(8)}} 
\startdata
58162.1319 &       2600.0 &       3 &       40 &       11.2 &     0.0043 &      <0.25&      <0.20\\
58164.859 &       1958.0 &       0 &       32 &       5.8 &     0.0030 &      <0.17&      <0.14\\
58166.5909 &       1728.0 &       1 &       23 &       7.8 &     0.0045 &      <0.26&      <0.21\\
58167.5201 &       2278.0 &       1 &       36 &       7.7 &     0.0034 &      <0.20&      <0.16\\
58170.8937 &       1221.0 &       1 &       15 &       7.9 &     0.0064 &      <0.37&      <0.30\\
58172.0895 &       1568.0 &       0 &       19 &       5.8 &     0.0037 &      <0.22&      <0.17\\
58766.7955 &       884.0 &       0 &       7 &       5.8 &     0.0066 &      <0.38&      <0.31\\
58831.6573 &       774.2 &       0 &       8 &       5.8 &     0.0075 &      <0.44&      <0.35\\
58834.0497 &       771.2 &       0 &       18 &       5.8 &     0.0075 &      <0.44&      <0.36\\
59794.4407 &       7821.0 &       3 &       74 &       10.8 &     0.0014 &      <0.08&      <0.07\\
\hline
58162-58172 &       11353.0 &       6 &       166 &       14.6 &     0.0013 &      <0.07&      <0.06\\
58831-58834 &       1545.4 &       0 &       26 &       5.8 &     0.0038 &      <0.22&      <0.18\\
\enddata
\tablecomments{ This table displays the results of the data reduction of the Swift/XRT observations. Column (2) lists the exposure time for each observation. Columns (3) and (4) show the number of counts in the source and background regions. Columns (5) and (6) provide the counts and upper limits of the source count rate. Columns (7) and (8) present the unabsorbed flux, with only the Galactic absorption taken into account, assuming a typical SED from TDEs or SNe.
}
\end{deluxetable}

\begin{deluxetable}{cc}
\setlength{\tabcolsep}{0.06in}
\tablecaption{ \it The SED of the host galaxy \label{host-photos} }
\tablewidth{0pt}
\tablehead{Filter & Flux  \\ 
                  & mJy    }
\startdata
2MASS H  &   11.100$\pm$0.607\\
2MASS J  &   10.100$\pm$0.427\\
2MASS Ks  &   9.300$\pm$0.748\\
PS1 g  &   2.306$\pm$0.017\\
PS1 i  &   4.969$\pm$0.016\\
PS1 r  &   3.850$\pm$0.041\\
PS1 y  &   6.590$\pm$0.105\\
PS1 z  &   5.700$\pm$0.046\\
WISE W1  &   5.230$\pm$0.039\\
WISE W2  &   3.130$\pm$0.052\\
WISE W3  &   14.900$\pm$0.233\\
WISE W4  &   24.000$\pm$1.610\\
GALEX FUV  &   0.164$\pm$0.014\\
GALEX NUV  &   0.249$\pm$0.010\\
SDSS g  &   2.371$\pm$0.009\\
SDSS i  &   5.529$\pm$0.024\\
SDSS r  &   4.085$\pm$0.018\\
SDSS u  &   0.764$\pm$0.016\\
SDSS z  &   6.608$\pm$0.062\\
\enddata
\tablecomments{The multi-band photometry of the host galaxy of ASASSN-18ap is listed in this table. }
\end{deluxetable}

\clearpage

\section{MOSFiT Fitting Results} 
\label{appendiceE}
\begin{figure*}[htb]
\centering
\begin{minipage}{0.9\textwidth}
\centering{\includegraphics[angle=0,width=1.0\textwidth]{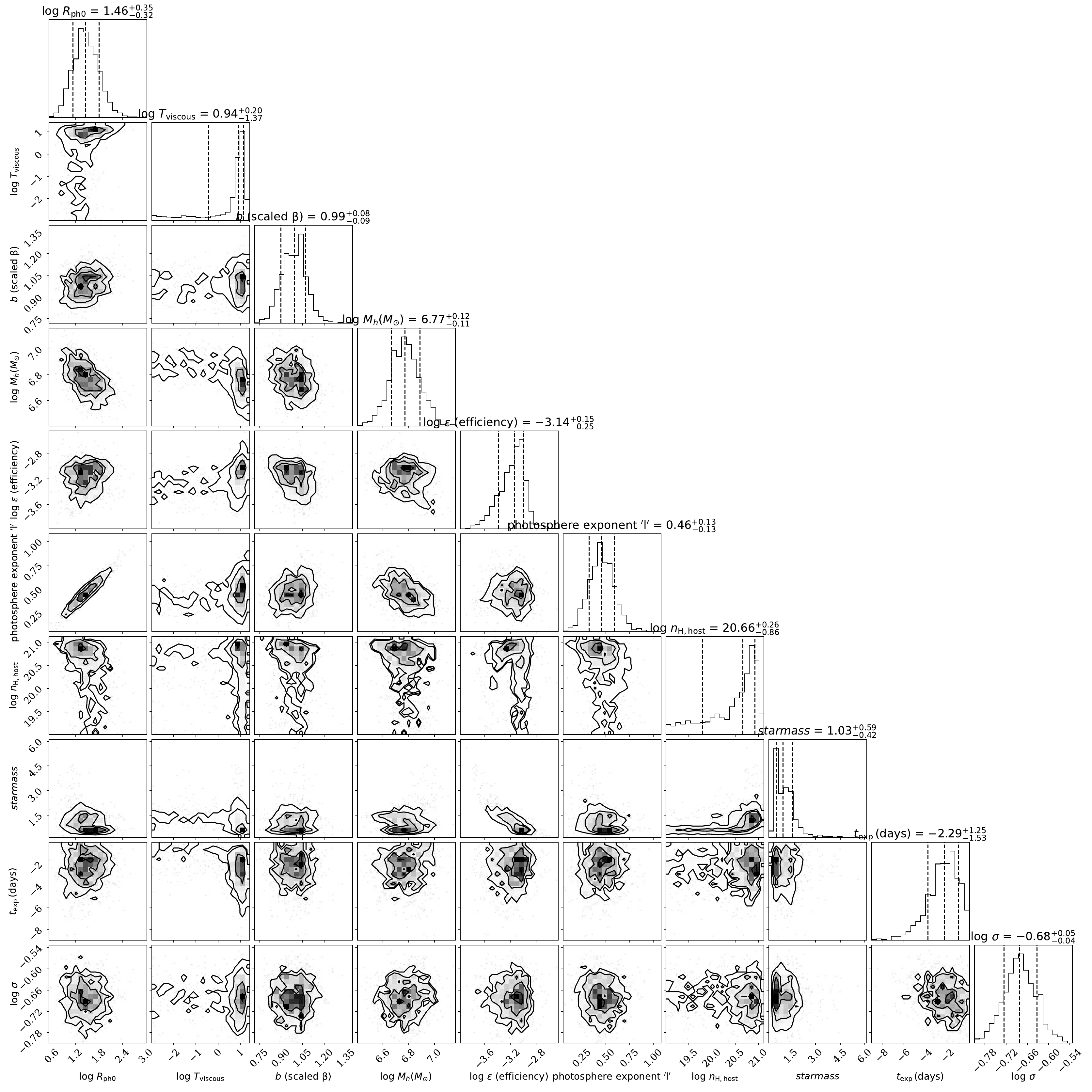}}
\end{minipage}
\caption{ We present the corner plots from the MOSFiT fitting to the multiwavelength light curve with the TDE model as an example. The corner plots of all three models are available in Figure set \ref{figset} in the online journal. \label{Cornerplot}}
\end{figure*}

In this appendix, we display the prior and posterior distribution of the MOSFiT parameters for models employed in Section \ref{sec3.5} in Table \ref{MOSFiT}, and the corner plots in Figset \ref{figset} in the online journal (refer to Figure \ref{Cornerplot} for an example).

\begin{deluxetable*}{cccc}
\setlength{\tabcolsep}{0.06in}
\tablecaption{\it Prior and posterior distribution of the MOSFiT parameters\label{MOSFiT}}
\tablewidth{0pt}
\tablehead{Parameters       &  Priors            &    Posteriors                & Units \\
           \colhead{(1)}    & \colhead{(2)}      &   \colhead{(3)}              & \colhead{(4)}}
\startdata
                            & \textbf{TDE model} &                              & \\
\hline
   star mass $\rm M_*$      & [0.01, 20]         &    $1.03^{+0.59}_{-0.42}$    & $\rm M_\odot$ \\ 
   $\rm log(M_{BH})$        & [5.0, 7.7]         &    $6.77^{+0.12}_{-0.11}$    & $\rm M_\odot$ \\
   b  (scaled $\beta$)      & [0.0, 2.0]         &    $0.99^{+0.08}_{-0.09}$    &               \\
   $\rm log(\epsilon)$      & [-5, -0.4]         &    $-3.14^{+0.15}_{-0.25}$   &               \\ 
   $\rm log(R_{ph0})$       & [-4., 4.0]         &    $1.46^{+0.35}_{-0.32}$    &               \\
   $\rm l_{ph}$             & [0.0, 4.0]         &    $0.46^{+0.13}_{-0.13}$    &               \\ 
   $\rm log(T_{viscous})$   & [-3.0,3.0]         &    $0.94^{+0.20}_{-1.37}$    &               \\  
   $\rm t_{exp}$            & [-50.0,0.0]        &    $-2.29^{+1.25}_{-1.53}$   & days          \\ 
   $\rm log(n_{H,Host})$    & [19.0, 23]         &    $20.66^{+0.26}_{-0.86}$   & $\rm cm^{-2}$ \\
   $\rm log(\sigma) $       & [-4.0,2.0]         &    $-0.68^{+0.05}_{-0.04}$   &               \\
\hline 
                            & \textbf{CSMNI model}&                             &               \\
\hline 
ejecta mass $\rm log(M_{ej})$ & [-1,2.477]        &   $1.83^{+0.14}_{-0.11}$     &$\rm M_\odot$  \\
CSM profile $\rm s$           & [0.0, 2.0]        &   $0.19^{+0.35}_{-0.14}$     &               \\
$\rm log(f_{Ni})$             & [-3.,-0.3]        &   $-1.40^{+0.13}_{-0.13}$    &               \\ 
$\rm t_{explosion}$           & [-200,0.0]        &   $-1.51^{+0.70}_{-1.23}$    & days              \\ 
$\rm log(T_{min})$            & [3.0, 5.0]        &   $3.72^{+0.04}_{-0.04}$     & K          \\ 
$\rm log(n_{H,Host})$         & [16 , 23 ]        &   $20.71^{+0.16}_{-0.66}$    & $\rm cm^{-2}$ \\
$\rm log(\kappa_\gamma)$      & [-1.0,4.0]        &   $1.61^{+1.67}_{-1.81}$     & $\rm cm^{-2}\,g^{-1}$ \\
$\rm log(M_{CSM})$            & [-1.0,2.0]        &   $-0.21^{+0.11}_{-0.12}$    & $\rm M_\odot$ \\ 
$\rm log(\rho)$               & [-15,-6.0]        &   $-10.61^{+0.63}_{-0.32}$   & $\rm cm^{-3}$     \\ 
$\rm log(\sigma)$             & [-5.0,1.0]        &   $-0.72^{+0.05}_{-0.05}$    &        \\
$\rm log(R_0)$                & [-1.0,3.0]        &   $0.07^{+0.72}_{-0.73}$     &          \\
\enddata

\tablecomments{ The prior and posterior of MOSFiT parameters. Column (1): The name of the parameter, with detailed definitions available in \citealt{Mockler2019} for TDE model, and \citealt{Villar2017} for CSMNI model. Column (2): The prior distribution of parameters, which is either uniform or log-uniform as indicated by their names in column (1). Column (3): The best fit value with 1-sigma error, and we didn't include the systematic error. Column (4): The units of the listed parameters. }
\end{deluxetable*}

\figsetstart
\figsetnum{1} 
\figsettitle{\label{figset} MOSFiT corner plots for the TDE,CSMNI modes}

\figsetgrpstart
\figsetgrpnum{figurenumber.1} 
\figsetgrptitle{TDE}
\figsetplot{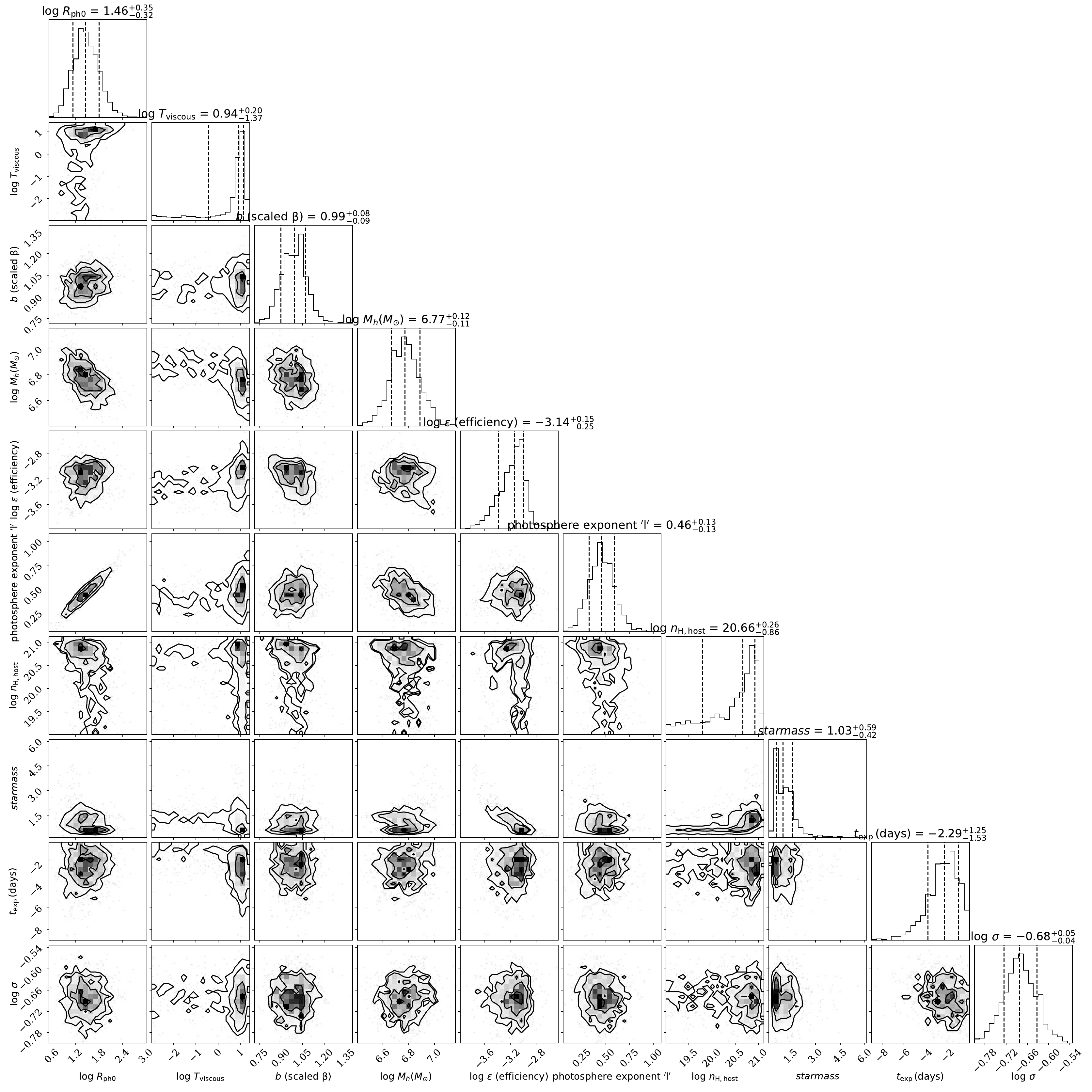} 
\figsetgrpnote{Corner plots from MOSFiT fitting to multiwavelength light curve with TDE model.}
\figsetgrpend

\figsetgrpstart 
\figsetgrpnum{figurenumber.2} 
\figsetgrptitle{CSMNI}
\figsetplot{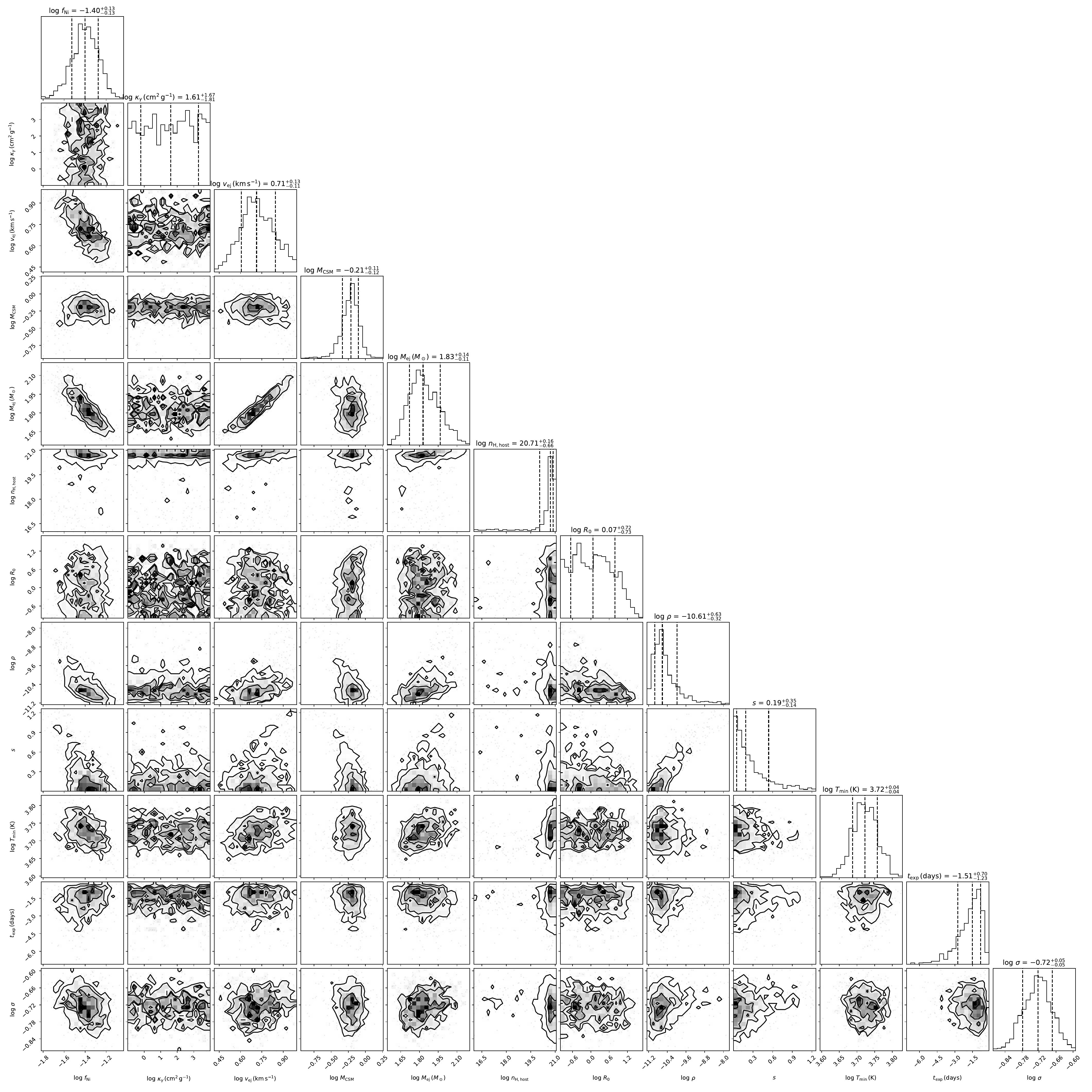} 
\figsetgrpnote{Corner plots from MOSFiT fitting to multiwavelength light curve with CSMNI model.}
\figsetgrpend

\figsetend 

\end{document}